\documentclass[letterpaper,11pt,onecolumn,titlepage]{article}

\usepackage[latin1]{inputenc}
\usepackage{amssymb, amsmath}
\usepackage[pdftex]{graphicx}
\usepackage{times}
\usepackage{subfig}
\usepackage{paralist}
\usepackage[compact]{titlesec}
\usepackage{boxedminipage}
\usepackage[pdftex]{hyperref}
\usepackage{listings}
\usepackage{fixltx2e} % contains fix for "1-col fig can come before earlier 2-col fig"

\renewenvironment{thebibliography}[1]{%
  \begin{oldthebibliography}{#1}%
    \setlength{\parskip}{0ex}%
    \setlength{\itemsep}{0ex}%
}%
{%
  \end{oldthebibliography}%
}

\begin{document}

\title{SEPIA: Security through Private Information Aggregation}

\author{
  Martin Burkhart, Mario Strasser, Dilip Many, Xenofontas Dimitropoulos \\ 
  \{burkhart, strasser, dmany, fontas\}@tik.ee.ethz.ch \\
  $\;$ \\
  Computer Engineering and Networks Laboratory,  ETH Zurich, Switzerland  \\
  $\;$ \\
  \textbf{TIK-Report No. 298}
}

\date{February 2009 \\ (updated February 2010)}

\maketitle

\begin{abstract}
Secure multiparty computation (MPC) allows joint privacy-preserving computations on data of 
multiple parties. Although MPC has been studied substantially,
building solutions that are practical in terms of computation and communication cost is
still a major challenge.
In this paper, we investigate the practical usefulness of MPC
for multi-domain network security and monitoring. 
We first optimize MPC comparison operations for processing high volume data in near real-time.
We then design privacy-preserving protocols for event correlation and
aggregation of network traffic statistics, such as addition of volume
metrics, computation of feature entropy, and distinct item count.   
Optimizing performance of parallel invocations, we implement our protocols
along with a complete set of basic operations in a library called SEPIA. We evaluate the running time and
bandwidth requirements of our protocols in realistic settings on a local cluster as well
as on PlanetLab and show
that they work in near real-time for up to 140 input
providers and 9 computation nodes. Compared to implementations using existing general-purpose
MPC frameworks, our protocols are significantly faster, requiring, for
example, 3 minutes for a task that takes~2 days with general-purpose
frameworks. This improvement paves the way for new applications of MPC in the area of networking.
Finally, we run SEPIA's protocols on real traffic traces of 
17 networks and show how they provide new possibilities for distributed troubleshooting and early
anomaly detection.
\end{abstract}

\section{Introduction}
\label{sec:intro}

\begin{figure}[t]
	\centering
	\includegraphics[scale=0.45]{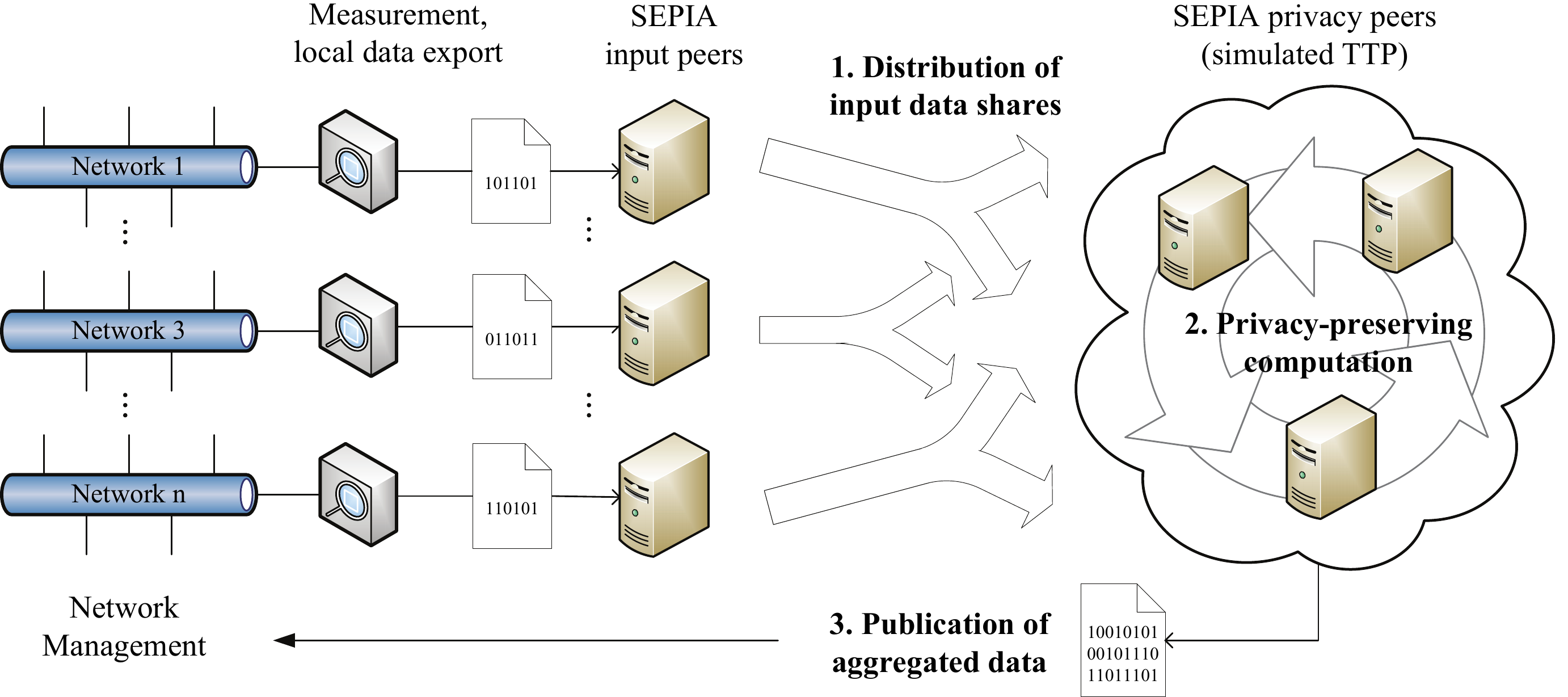}
	\caption{Deployment scenario for SEPIA.}
	\label{fig:scenario}
\end{figure}

A number of network security and monitoring problems can substantially
benefit if a group of involved organizations aggregates private data
to jointly perform a computation. For example, IDS alert correlation,
e.g., with DOMINO~\cite{yegneswaran2004gid}, requires the joint
analysis of private alerts. Similary, aggregation of private data is
useful for alert signature extraction~\cite{janak2006}, collaborative
anomaly detection~\cite{ringberg2009}, multi-domain traffic
engineering~\cite{Machi2004}, detecting traffic
discrimination~\cite{tariq2009}, and collecting network performance
statistics~\cite{Simpson2004}. All these approaches use either a
trusted third party, e.g., a University research group, or
peer-to-peer techniques for data aggregation and face a delicate
privacy versus utility trade-off~\cite{porras2006lcs}. 
Some private data typically have to be revealed, which impedes privacy and
prohibits the acquisition of many data providers, while data anonymization,
used to remove sensitive information, complicates or even prohibits
developing good solutions.  
Moreover, the ability of anonymization techniques to effectively protect privacy is questioned by recent
studies~\cite{Ohm2010}. One possible solution to this privacy-utility trade-off is MPC.

For almost thirty years, MPC~\cite{yao1982psc} techniques have been studied for
solving the problem of jointly running computations on data
distributed among multiple organizations, while provably preserving
data privacy without relying on a trusted third party. In theory, any
computable function on a distributed dataset is also securely
computable using MPC techniques~\cite{Goldreich1987any}. However, designing 
solutions that are practical in terms of running time and
communication overhead is non-trivial. For this reason, MPC techniques have mainly
attracted theoretical interest in the last decades. Recently,
optimized basic primitives, such as
comparisons~\cite{damgard2006bitdecomp,nishide2007nobitdecomp}, make
progressively possible the use of MPC in real-world applications,
e.g., an actual sugar-beet auction~\cite{bogetoft2009secure} was
demonstrated in 2009.

Adopting MPC techniques to network monitoring and security problems
introduces the important challenge of dealing with voluminous input
data that require online processing. For example, anomaly detection
techniques typically require the online generation of traffic volume
and distributions over port numbers or IP address ranges. Such input
data impose stricter requirements on the performance of MPC protocols
than, for example, the input bids of a distributed MPC
auction~\cite{bogetoft2009secure}.  In particular, network monitoring
protocols should process potentially thousands of input values while
meeting {\it near real-time} guarantees\footnote{We define {\it near
real-time} as the requirement of fully processing an $x$-minute
interval of traffic data in no longer than $x$ minutes, where $x$ is typically
a small constant. For our evaluation, we use 5-minute windows.}. This is not
presently possible with existing general-purpose MPC frameworks.

In this work, we design, implement, and evaluate SEPIA (Security
through Private Information Aggregation), a library for
efficiently aggregating multi-domain network data using MPC. 
The foundation of SEPIA is a set of optimized MPC operations, implemented
with performance of parallel execution in mind.
By not enforcing protocols to run in a constant number of rounds, 
we are able to design MPC comparison operations that require up to 80 times
less distributed multiplications and, amortized over many parallel invocations, 
run much faster than constant-round alternatives. 
On top of these comparison operations, we
design and implement novel MPC protocols tailored for network
security and monitoring applications. The {\it event correlation}
protocol identifies events, such as IDS or firewall alerts, that occur
frequently in multiple domains.  The protocol is generic having
several applications, for example, in alert correlation for early
exploit detection or in identification of multi-domain network traffic
heavy-hitters. In addition, we introduce SEPIA's {\it entropy} and
{\it distinct count} protocols that compute the entropy of traffic
feature distributions and find the count of distinct feature values,
respectively. These metrics are used frequently in traffic analysis
applications.  In particular, the entropy of feature distributions is
used commonly in anomaly detection, whereas distinct count metrics are
important for identifying scanning attacks, in firewalls, and for
anomaly detection. We implement these protocols along with a
vector addition protocol to support additive
operations on timeseries and histograms.

A typical setup for SEPIA is depicted in Fig.~\ref{fig:scenario} where individual networks
are represented by one \emph{input peer} each. The input peers distribute shares of secret
input data among a (usually smaller) set of \emph{privacy peer}s using Shamir's secret sharing scheme~\cite{shamir1979ss}.
The privacy peers perform the actual computation and can be hosted by a subset of the networks running input peers but also 
by external parties. Finally, the aggregate computation result is sent back to the networks.
We adopt the semi-honest adversary model, hence privacy of local input data is guaranteed as long as no more than half of the privacy peers collude.

Our evaluation of SEPIA's performance shows that SEPIA runs in near 
real-time even with~140 input and 9 privacy peers. Moreover, we run SEPIA on traffic data of~17
networks collected during the global Skype outage in August 2007 and
show how the networks can use SEPIA to troubleshoot and
timely detect such anomalies. Finally, we discuss novel applications
in network security and monitoring that SEPIA enables.  In summary,
this paper makes the following contributions:

\vspace{1mm}
\begin{compactenum}

\item We introduce efficient MPC comparison operations, which
outperform constant-round alternatives for many parallel invocations.

\item We design novel MPC protocols for event correlation, entropy and
distinct count computation.

\item We introduce the SEPIA library, in which we implement our
protocols along with a complete set of basic operations, optimized for
parallel execution. SEPIA is made publicly available.

\item We extensively evaluate the performance of SEPIA on realistic
settings using synthetic and real traces and show that it
meets near real-time guarantees even with~140 input and 9
privacy peers.

\item We run SEPIA on traffic from~17 networks and show how it
can be used to troubleshoot and timely detect anomalies, exemplified by the Skype outage. 
%Moreover, we outline other applications in network security and
%monitoring that SEPIA enables.

\end{compactenum}
\vspace{1mm}

The paper is organized as follows: We specify the computation scheme
in the next section and present our optimized comparison operations in
Section~\ref{sec:comparisons}.  In Section~\ref{sec:protocols}, we
build the protocols for event correlation, vector addition, entropy
and distinct count computation.  We evaluate the protocols and discuss
SEPIA's design in Sections~\ref{sec:evaluation} and~\ref{sec:design},
respectively. Then, in Section~\ref{sec:applications} we outline
SEPIA's applications and conduct a case study on real network data
that demonstrates SEPIA's benefits in distributed troubleshooting and
early anomaly detection. Finally, we discuss related work in
Section~\ref{related} and conclude our paper in
Section~\ref{sec:conclusion}.

\section{Preliminaries}
\label{sec:preliminaries}

Our implementation is based on Shamir secret
sharing~\cite{shamir1979ss}.  In order to \emph{share} a secret value $s$
among a set of $m$ players, the dealer generates a random
polynomial~$f$ of degree~$t=\left\lfloor (m-1)/2 \right\rfloor$ over
a prime field $\mathbb{Z}_p$ with $p>s$, such that $f(0)=s$.  
Each player $i=1\ldots m$ then receives an evaluation
point $s_i=f(i)$ of $f$. $s_i$ is called the share of player $i$.
The secret $s$ can be reconstructed from any~$t+1$
shares using Lagrange interpolation but is completely undefined for
$t$ or less shares. To actually \emph{reconstruct} a secret, each 
player sends his shares to all other players. Each player
then locally interpolates the secret.
For simplicity of presentation, we 
use $[s]$ to denote the vector of shares $(s_1, \ldots, s_m)$ and call it a \emph{sharing} of $s$. 
In addition, we use $[s]_i$ to refer to $s_i$.
Unless stated otherwise, we choose $p$ with 62 bits such that arithmetic operations
on secrets and shares can be performed by CPU instructions directly,
not requiring software algorithms to handle big integers.  

\paragraph{Addition and Multiplication} 

Given two sharings $[a]$ and $[b]$, we can perform private addition and multiplication of the two values $a$ and $b$. 
Because Shamir's scheme is linear, addition of two sharings, denoted by $[a]+[b]$, can be computed by having 
each player locally add his shares of the two values: $[a+b]_i=[a]_i+[b]_i$.
Similarly, local shares are subtracted to get a share of the difference.
To add a public constant $c$ to a sharing $[a]$, denoted by $[a]+c$, each player just adds $c$ to his share, i.e.,
$[a+c]_i = [a]_i+c$. Similarly, for multiplying $[a]$ by a public constant $c$, denoted by $c[a]$, each player multiplies its share by $c$.
Multiplication of two sharings requires an extra round of communication to guarantee randomness and to correct the degree of the
new polynomial~\cite{benor1988ctn,gennaro1998sva}. In particular, to compute $[a][b]=[ab]$, each player first computes $d_i=[a]_i[b]_i$ locally. He
then shares $d_i$ to get $[d_i]$. Together, the players then perform a distributed Lagrange interpolation to compute $[ab] = \sum_i{\lambda_i[d_i]}$ where $\lambda_i$ are the Lagrange coefficients. Thus, a distributed multiplication requires a synchronization round with $m^2$ messages, as each player $i$ sends to each player $j$ the share $[d_i]_j$. 
To specify protocols, composed of basic operations, we use a shorthand notation. For instance, 
we write $foo([a],b):=([a]+b)([a]+b)$, where $foo$ is the protocol name, followed by input parameters. Valid input parameters are sharings and public
constants. On the right side, the function to be computed is given, a binomial in that case. The output of $foo$ is again a sharing and can be used in 
subsequent computations.
All operations in $\mathbb{Z}_p$ are performed modulo $p$,
therefore $p$ must be large enough to avoid modular reductions of
intermediate results, e.g., if we compute $[ab]=[a][b]$, then $a$, $b$, and $ab$ must be smaller than $p$.

\paragraph{Communication}
A set of independent multiplications, e.g., $[ab]$ and $[cd]$, can be performed
in parallel in a single round. That is, intermediate results of all multiplications
are exchanged in a single synchronization step. \emph{A round simply is a
synchronization point where players have to exchange intermediate results} in order
to continue computation. 
While the specification of the protocols is synchronous, we do not assume the network
to be synchronous during runtime. In particular, the Internet is better modeled as asynchronous, not guaranteeing
the delivery of a message before a certain time. Because we assume the semi-honest model, we only have to
protect against high delays of individual messages, potentially leading to a reordering of message arrival.
In practice, we implement communication channels using SSL sockets over TCP/IP. TCP applies
acknowledgments, timeouts, and sequence numbers to preserve message ordering and to retransmit
lost messages, providing FIFO channel semantics. We implement message synchronization in parallel threads to minimize waiting time.
Each player proceeds to the next round immediately after sending and receiving all intermediate values.

%We consider the sending and the receiving of messages as
%asynchronous events, but assume that there exists a constant upper bound on the
%delivery time of messages. Messages which are not delivered during this time
%are considered as lost. Also, messages arrive in the same order they were sent.

%\footnote{The protocol operation parallelism in one round is not to be confused with parallelization of local CPU operations. Although we may execute many parallel multiplications in a single round, the local CPU cycles required for performing these multiplications are in general still sequential. Using multiple CPUs/cores to locally solve these multiplications is an orthogonal optimization step. SEPIA, for example, parallelizes local CPU tasks by using several threads (see Section~\ref{sec:design}).}.

\paragraph{Security Properties} 
All the protocols we devise are compositions of the above introduced
addition and multiplication primitives, which were proven correct and
\emph{information-theoretically} secure by Ben-Or, Goldwasser, and
Wigderson~\cite{benor1988ctn}. In particular, they showed that in the
semi-honest model, where players follow the protocol but try to learn
as much as possible by sharing the information they received, no set
of $t$ or less players gets any additional information other than the
final function value.  Also, these primitives are~\emph{universally
composable}, that is, the security properties remain intact under
stand-alone and concurrent composition~\cite{canetti2001universally}.

%For convenience, we list the notation and variables used in Table~\ref{tab:notation}.
%
%\begin{table}[t]
%	\centering
%		\begin{tabular}{rlrlrl}
%			$n$: 	& number of input peers 			& $m$:		& number of privacy peers 		& $p$: & prime field size \\
%			$l$: 	&size of shares in bits ($l=\log_2 p)$ & $[b]$: & value $b$ shared as a secret 			& $[b]_i$: & the $i$-th share of $[b]$ 
%		\end{tabular}
%		\caption{Overview of notation and important variables.}
%		\label{tab:notation}
%\end{table}

\section{Optimized Comparison Operations} 
\label{sec:comparisons}
Unlike addition and multiplication, comparison of two shared secrets is a very expensive operation. Therefore, we now devise optimized protocols for
equality check, less-than comparison and a short range check. 
The complexity of an MPC protocol is typically assessed counting the number of distributed multiplications and rounds, because addition and multiplication with public values only require local computation. 
Damg{\aa}rd \emph{et al.} introduced the bit-decomposition protocol~\cite{damgard2006bitdecomp} 
that achieves comparison by decomposing shared secrets into a shared bit-wise representation. On shares
of individual bits, comparison is straight-forward. With \mbox{$l=\log_2(p)$}, the protocols in~\cite{damgard2006bitdecomp} achieve a comparison with $205l+188l\log_{2}l$ multiplications in 44 rounds and equality test with $98l+94l\log_{2}l$ multiplications in 39 rounds. 
Subsequently,~\cite{nishide2007nobitdecomp} have improved these protocols by not decomposing the secrets but using bitwise shared random numbers. They 
do comparison with $279l+5$ multiplications in 15 rounds and equality test with $81l$ multiplications in 8 rounds.
While these are constant-round protocols as preferred in theoretical research, they still involve lots of multiplications. For instance, an equality check of two shared IPv4 addresses ($l=32$) with~\cite{nishide2007nobitdecomp} requires $2592$ distributed multiplications, each triggering $m^2$ messages to be transmitted over the network.
%In this Section, we design comparison operations requiring significantly less multiplications.

\paragraph{Constant-round vs. number of multiplications}
Our key observation for improving efficiency is the following: For scenarios with many parallel protocol invocations it is 
possible to build much more practical protocols by not enforcing the constant-round property.
Constant-round means that the number of rounds does not depend on the input parameters. We design protocols that run in O($l$) rounds and are therefore not constant-round, although, once the field size $p$ is defined, the number of rounds is also fixed, i.e., not varying at runtime.
The overall local running time of a protocol is determined by i) the local CPU time spent on computations, ii) the time to transfer intermediate values over the network, and iii) delay experienced during synchronization.
Designing constant-round protocols aims at reducing the impact of iii) by keeping the number of rounds fixed and usually small. To achieve this, high multiplicative constants for the number of multiplications are often accepted (e.g., $279l$). Yet, both i) and ii) directly depend on the number of multiplications. 
For applications with few parallel operations, protocols with few rounds (usually constant-round) are certainly faster. However, with many parallel operations, as required by our scenarios, the impact of network delay is amortized and the number of multiplications (the actual workload) becomes the dominating factor. Our evaluation results in Section~\ref{sec:eval_event} and ~\ref{sec:eval_frameworks} confirm this and show that CPU time and network bandwidth are the main constraining factors, calling for a reduction of multiplications.

%Therefore, we design and use comparison protocols that significantly reduce the number of multiplications by trading it off against the constant-round property.

%Albeit not as radical, similar tradeoffs are made by other practical applications of comparison operations. For instance, the version by Toft~\cite{toft05Comparison} uses $log(l)$ rounds and is implemented as the standard comparison operation in VIFF~\cite{VIFF}. 
%For a direct performance comparison with VIFF please refer to Section~\ref{sec:eval_frameworks}.

\paragraph{Equality Test} 
In the field $\mathbb{Z}_p$ with $p$ prime, Fermat's little theorem states
\begin{equation}
c^{p-1} = 
\begin{cases} 0 & \text{if $c=0$} \\
							1 &\text{if $c \neq 0$}
\end{cases}
\label{eqn:flt}
\end{equation}
Using~\eqref{eqn:flt} we define a protocol for equality test as follows:
\begin{equation*}
equal([a],[b]) := 1-([a]-[b])^{p-1}
\label{eqn:equal}
\end{equation*}
The output of $equal$ is $[1]$ in case of equality and $[0]$ otherwise and can hence be used in subsequent computations. Using square-and-multiply for the exponentiation, we implement $equal$ with $l+k-2$ multiplications in $l$ rounds, where $k$ denotes the number of bits set to $1$ in $p-1$. By using carefully picked prime numbers with $k \leq 3$, we reduce the number of multiplications to $l+1$. In the above example for comparing IPv4 addresses, this reduces the multiplication count by a factor of $76$ from $2592$ to $34$.
%This reduction is essential because it allows us to perform thousands of comparisons per second, and in turn meet the near real-time requirements for our event correlation protocol (see Section~\ref{sec:evaluation}).

Besides having few 1-bits, $p$ must be bigger than the range of shared secrets, i.e., if 32-bit integers are shared, an appropriate $p$ will have at least 33 bits. For any secret size below 64 bits it is easy to find appropriate $p$s with $k \leq 3$ within 3 additional bits.

\paragraph{Less Than}

%\begin{figure}[t]
	%\centering
	%\includegraphics[clip,angle=270,scale=0.4]{figures/operations} 
	%\caption{Supported operations and subprotocols. Bold 
%operations are used directly in SEPIA protocols.}
	%\label{fig:operations}
%\end{figure}

For less-than comparison, we base our implementation on Nishide's protocol~\cite{nishide2007nobitdecomp}. However, we apply modifications to again reduce the overall number of required multiplications by more than a factor of 10. Nishide's protocol is quite comprehensive and built on a stack of subprotocols for 
least-significant bit extraction (LSB), operations on bitwise-shared secrets, and (bitwise) random number sharing. The protocol uses the observation that $a<b$ is determined by the three predicates $a<p/2$, $b<p/2$, and $a-b<p/2$. Each predicate is computed by a call of the LSB protocol for $2a$, $2b$, and $2(a-b)$. If $a<p/2$, no wrap-around modulo $p$ occurs when computing $2a$, hence $LSB(2a)=0$. However, if $a>p/2$, a wrap-around will occur and $LSB(2a)=1$.
Knowing one of the predicates in advance, e.g., because $b$ is not secret but publicly known, saves one of the three LSB calls and hence $1/3$ of the multiplications. 

Due to space restrictions we omit to reproduce the entire protocol but focus on the modifications we apply. An important subprotocol in Nishide's construction is $PrefixOr$. Given a sequence of shared bits $[a_1],\ldots,[a_l]$ with $a_i \in \{0,1\}$, $PrefixOr$ computes the sequence $[b_1],\ldots,[b_l]$ such that $b_i = \vee_{j=1}^{i} a_j$. Nishide's $PrefixOr$ requires only 7 rounds but $17l$ multiplications. We implement $PrefixOr$ based on the fact that $b_i = b_{i-1} \vee a_i$ and $b_1=a_1$. The logical \texttt{OR} ($\vee$) can be computed using a single multiplication: $[x] \vee [y] = [x]+[y]-[x][y]$. Thus, our $PrefixOr$ requires $l-1$ rounds and only $l-1$ multiplications. 

Without compromising security properties, we replace the $PrefixOr$ in Nishide's protocol by our optimized version and call the resulting comparison protocol $lessThan$. A call of $lessThan([a],[b])$ outputs $[1]$ if $a<b$ and [0] otherwise. The overall complexity of $lessThan$ is $24l+5$ multiplications in $2l+10$ rounds as compared to Nishide's version with $279l+5$ multiplications in $15$ rounds. 

\paragraph{Short Range Check}
To further reduce multiplications for comparing small numbers, we devise a check for short ranges, based on our $equal$ operation.
Consider one wanted to compute $[a]<T$, where T is a small public constant, e.g., $T=10$. Instead of invoking $lessThan([a],T)$ one can simply compute the polynomial
$[\phi]=[a]([a]-1)([a]-2)\ldots([a]-(T-1))$. If the value of $a$ is between $0$ and $T-1$, exactly one term of $[\phi]$ will be zero and hence $[\phi]$
will evaluate to [0]. Otherwise, $[\phi]$ will be non-zero. Based on this, we define a protocol for checking short public ranges that returns $[1]$ if $x \leq [a] \leq y$ and $[0]$ otherwise:
\begin{equation*}
shortRange([a], x, y) := equal \bigl(0, \prod_{i=x}^{y}{([a]-i)} \bigr)
\end{equation*}
The complexity of $shortRange$ is $(y-x)+l+k-2$ multiplications in $l+\log_2 (y-x)$ rounds. 
Computing $lessThan([a],y)$ requires $16l+5$ multiplications (1/3 is saved because $y$ is public).
Hence, regarding the number of multiplications, computing $shortRange([a],0,y-1)$ instead of $lessThan([a],y)$ is beneficial roughly as long as $y \leq 15l$.

\section{SEPIA Protocols}
\label{sec:protocols}

In this section, we compose the basic operations defined above into full-blown protocols for network event correlation and statistics aggregation.
We first define the basic setting of SEPIA protocols as illustrated in Fig.~\ref{fig:scenario} and then introduce the protocols successively.

Our system has a set of $n$ users called \emph{input peers}. The input peers want to jointly compute the value of a public function~$f(x_1,\ldots,x_n)$ on their private data $x_i$ without disclosing anything about $x_i$.
In addition, we have $m$ players called \emph{privacy peers} that perform the computation of~$f()$ by simulating
a trusted third party (TTP). Each entity can take both roles, acting only as an input peer, privacy peer (PP) or both.
We use the semi-honest (a.k.a. honest-but-curious) adversary model for privacy peers. That is, adversarial
privacy peers do follow the protocol but try to infer as much as possible from the values (shares) they learn. 
The privacy and correctness guarantees provided by our protocols are determined by Shamir's secret
sharing scheme.
The protocols are secure against $t < m/2$ colluding privacy peers. That is, in order to protect against at least one curious privacy peer, $m$ has to be larger than 2.

The function $f()$ is specified \emph{as if} a TTP was available. The MPC scheme then guarantees that
no information is leaked \emph{from the computation process}. However, just learning the resulting value 
$f()$ could allow to deduce sensitive information.
For example, if the input bit of all input peers must remain secret, computing the logical \texttt{AND} of all
input bits is insecure in itself: if the final result was $1$, all input bits must be $1$ as well and are thus no longer secret.
\emph{It is the responsibility of the input peers to verify that learning $f()$ is acceptable}, 
in the same way as they have to verify this when using a real TTP.
For example, in our protocols we assume input peers are not willing to reconstruct complete item distributions but consider it
safe to compute the overall item count or entropy.
To reduce the potential for deducing information from $f()$, protocols can enforce the submission of ``valid'' input data.
For instance, in our event correlation protocol, the privacy peers verify that each input peer submits no duplicate events. 

Note that although the number of privacy peers $m$ has a quadratic
impact on the total communication and computation costs, there are also~$m$ privacy
peers sharing the load. That is, if the network capacity is sufficient, the overall running time of the
protocols will scale linearly with~$m$ rather than quadratically.
On the other hand, the number of tolerated colluding privacy peers
also scales linearly with $m$. Hence, the choice of $m$ involves a privacy-performance tradeoff. The separation of
roles into input and privacy peers allows to tune this tradeoff independently of the number of input providers.

Prior to running the protocols, the $m$ privacy peers set up a secure,
i.e., confidential and authentic, channel to each other. In
addition, each input peer creates a secure channel to each
privacy peer. We assume that the required public keys and/or
certificates have been securely distributed beforehand. 
%As we do not want to engage in delicate tradeoffs, \emph{our goal is to develop protocols that do not leak sensitive intermediate information}.
All protocols are designed to run on continuous streams of input traffic data partitioned into time windows of a few minutes. In the following,
each protocol is specified for a single time window.

\subsection{Event Correlation}
\label{sec:ec}

\begin{figure*}[t]
	\begin{boxedminipage}{\textwidth}
		\begin{small}
		\begin{compactenum}
		
		\item \textbf{Share Generation:} Each input peer~$i$ shares $s$ distinct events $e_{ij}$ with $w_{ij} < w_{max}$ among the privacy peers (PPs).
		
		\item \textbf{Weight Verification:} Optionally, the PPs compute and reconstruct $lessThan([w_{ij}], w_{max})$ for all weights to verify that they are smaller  than $w_{max}$. Misbehaving input peers are disqualified.
		
		\item \textbf{Key Verification:} Optionally, the PPs verify that each input peer $i$ reports distinct events, i.e., for each event index $a$ and $b$ with $a<b$ they compute and reconstruct $equal([k_{ia}], [k_{ib}])$. Misbehaving input peers are disqualified.
		
		\item \textbf{Aggregation:} The PPs compute $[C_{ij}]$ and $[W_{ij}]$ according to~\eqref{eqn:cw} for $i \leq \hat{i}$ with $\hat{i}=min(n-T_c+1,n)$.~\footnote{For instance, if $n=10$ and $T_c=7$, each event that needs to be reconstructed according to~\eqref{eqn:cond} must be reported by at least one of the first 4 input peers. Hence, it is sufficient to compute the $C_{ij}$ and $W_{ij}$ for the first $n-T_c+1=4$ input peers.} All required $equal$ operations can be performed in parallel.
		\item \textbf{Reconstruction:} For each event $[e_{ij}]$, with $i \leq \hat{i}$, condition~\eqref{eqn:cond} has to be checked. Therefore, the PPs compute
		\begin{equation*}
		  {[t_1]} = shortRange([C_{ij}], T_c, n), \;\;\;\;\; {[t_2]} = lessThan(T_w-1, [W_{ij}])   
		\end{equation*}
		Then, the event is reconstructed iff $[t_1] \cdot [t_2]$ returns $1$. 
		The set of input peers with $i > \hat{i}$ reporting a reconstructed event $\overline{r}=(\overline{k},\overline{w})$ is computed by reusing all the $equal$ operations performed on $\overline{r}$ in the aggregation step. That is, input peer $i'$ reports $\overline{r}$ iff $\sum_{j}{equal([\overline{k}],[k_{i'j}])}$ equals 1. This can be computed using local addition for each remaining input peer and each reconstructed event. Finally, all reconstructed events are sent to all input peers.
		\end{compactenum}
		\end{small}
	\end{boxedminipage}
	\caption{Algorithm for event correlation protocol.}
	\label{box:eventcorrelation}
\end{figure*}

The first protocol we present enables the input peers to privately aggregate arbitrary network events.
An event $e$ is defined by a key-weight pair $e=(k, w)$. This notion is generic in the sense that keys can be defined to represent arbitrary types of network events, which are uniquely identifiable. The key $k$ could for instance be the source IP address of packets triggering IDS alerts, or the source address concatenated with a specific alert type or port number. It could also be the hash value of extracted malicious payload or represent a uniquely identifiable object, such as popular URLs, of which the input peers want to compute the total number of hits. The weight $w$ reflects the impact (count) of this event (object), e.g., the frequency of the event in the current time window or a classification on a severity scale.

Each input peer shares at most $s$ local events per time window. The goal of the protocol is to reconstruct an event if and only if a minimum number of input peers $T_c$ report the same event and the aggregated weight is at least $T_w$. The rationale behind this definition is that an input peer does not want to reconstruct local events that are unique in the set of all input peers, exposing sensitive information asymmetrically. But if
the input peer knew that, for example, three other input peers report the same event, e.g., a specific intrusion alert, he would be willing to contribute his information
and collaborate. Likewise, an input peer might only be interested in reconstructing events of a certain impact, having a non-negligible aggregated weight. 
%By setting any of the thresholds to $1$, the input peers could of course reconstruct as soon as only one of the two conditions is met.

More formally, let $[e_{ij}]=([k_{ij}], [w_{ij}])$ be the shared event $j$ of input peer $i$ with $j \leq s$ and $i \leq n$. Then we compute the aggregated count $C_{ij}$ and weight $W_{ij}$ according to \eqref{eqn:cw} and reconstruct $e_{ij}$ iff \eqref{eqn:cond} holds.

\begin{align}
[C_{ij}] &:= \sum_{i' \neq i, j'}{equal([k_{ij}], [k_{i'j'}])}  &  [W_{ij}] &:= \sum_{i' \neq i, j'}{[w_{i'j'}] \cdot equal([k_{ij}], [k_{i'j'}])} \label{eqn:cw}
\end{align}
\begin{equation}
	([C_{ij}] \geq T_c) \wedge ([W_{ij}] \geq T_w) \label{eqn:cond}
\end{equation}
%\begin{eqnarray}
%	[C_{ij}] & := & \sum_{i' \neq i, j'}{equal([k_{ij}], [k_{i'j'}])}  \label{eqn:c} \\
%	\relax[W_{ij}] & := & \sum_{i' \neq i, j'}{[w_{i'j'}] \cdot equal([k_{ij}], [k_{i'j'}])} \label{eqn:w}
%\end{eqnarray}

Reconstruction of an event $e_{ij}$ includes the reconstruction of $k_{ij}$, $C_{ij}$, $W_{ij}$, and the list of input peers reporting it, but the $w_{ij}$ remain secret. The detailed algorithm is given in Fig.~\ref{box:eventcorrelation}.

\paragraph{Input Verification} 

In addition to merely implementing the correlation logic, we devise two optional input verification steps. In particular the PPs check that shared weights are below a maximum weight $w_{max}$ and that each input peer shares distinct events. These verifications serve two purposes. First, they protect from misconfigured input peers and flawed input data. Secondly, they protect against input peers that try to deduce information from the final computation result. For instance, 
an input peer could add an event $T_c-1$ times (with a total weight of at least $T_w$) to find out whether any other input peers report the same event. These input verifications mitigate such attacks.

\paragraph{Complexity.} The overall complexity, including verification steps, is summarized below in terms of operation invocations and rounds:

\vspace{2mm}
\begin{tabular}{llll}
	$equal$:						& $O \bigl((n-T_c)ns^2 \bigr)$  & $lessThan$:				  &	$(2n-T_c)s$		\\
	$shortRange$:				&	$(n-T_c)s$		&	multiplications:	  &	$(n-T_c) \cdot (ns^2+s)$		\\
	%\hline
	rounds:						  &	$7l+ \log_2(n-T_c) + 26$		
\end{tabular}
\vspace{2mm}

%           									  equals	  					lessThan    shortRange 	   Multiplications
%           									------------------------------------------------------------------
% 2. Weight Verification   	               						ns
% 3. Key Verification      		ns(s-1)/2
% 4. Aggregation(d=\hat{i})		s^2(dn - d/2 -d^2/2)																d*n*s^2
% 5. Reconstruction									ds								ds             ds						
%
% Overall: 

The protocol is clearly dominated by the number of $equal$ operations required for the aggregation step. It scales quadratically with $s$, however, depending on $T_c$, it scales linearly or quadratically with $n$. 
For instance, if $T_c$ has a constant offset to $n$ (e.g., $T_c = n-4$), only	$O(ns^2)$ $equal$s are required. However, if $T_c = n/2$, $O(n^2s^2)$ $equal$s
are necessary.

\paragraph{Optimizations} To avoid the quadratic dependency on $s$, we are working on an MPC-version of
a binary search algorithm that finds a secret $[a]$ in a sorted list of secrets $\{[b_1], \ldots, [b_s]\}$ with $\log_2 s$ comparisons by 
comparing $[a]$ to the element in the middle of the list, here called $[b_*]$. We then construct a new list, being
the first or second half of the original list, depending on $lessThan([a],[b_*])$. The procedure is repeated recursively until the list has size 1.
This allows us to compare all events of two input peers with only $O(s \log_2 s)$ instead of $O(s^2)$ comparisons.
To further reduce the number of $equal$ operations, the protocol can be adapted
to receive \emph{incremental updates} from input peers. That is, input peers submit a list of events in each time window and
inform the PPs, which event entries have a different key from the previous window. Then, only comparisons of updated keys have to be performed and overall complexity is reduced to $O(u(n-T_c)s)$, where $u$ is the number of changed keys in that window. This requires, of course, that information on input set dynamics is not considered private.

\subsection{Network Traffic Statistics}
\label{sec:statistics}

In this section, we present protocols for the computation of multi-domain traffic statistics including
the aggregation of additive traffic metrics, the computation of feature entropy, and the computation of
distinct item count. These statistics find various applications in network monitoring and management.

\begin{figure}[t]
	\begin{boxedminipage}{1\textwidth}
	\begin{small}
	\begin{compactenum}
		\item \textbf{Share Generation:} Each input peer~$i$ shares its input vector $\mathbf{d_i}=(x_1, x_2, \ldots, x_r)$ among the PPs. That is, the PPs obtain $n$ vectors of sharings $[\mathbf{d_i}] = ([x_1], [x_2], \ldots, [x_r])$.
		\item \textbf{Summation:} The PPs compute the sum $[\mathbf{D}] = \sum_{i=1}^{n}{[\mathbf{d_i}]}$.
		\item \textbf{Reconstruction:} The PPs reconstruct all elements of $\mathbf{D}$ and send them to all input peers.
	\end{compactenum}
	\end{small}
	\end{boxedminipage}
	\caption{Algorithm for vector addition protocol.}
	\label{box:addition}
\end{figure}

\subsubsection{Vector Addition}
\label{sec:addition}

To support basic additive functionality on timeseries and histograms, we implement a vector addition protocol.
Each input peer $i$ holds a private $r$-dimensional input vector $\mathbf{d_i} \in \mathbb{Z}_p^r$. 
Then, the vector addition protocol computes the sum \(\mathbf{D} = \sum_{i=1}^{n}\mathbf{d_i} \). We describe the 
corresponding SEPIA protocol shortly in Fig.~\ref{box:addition}. 
This protocol requires no distributed multiplications and only one round.

\subsubsection{Entropy Computation}
\label{sec:entropy}

\begin{figure}[t]
	\begin{boxedminipage}{1\textwidth}
	\begin{small}
	\begin{compactenum}
	
	\item \textbf{Share Generation:} Each input peer holds an $r$-dimensional private input vector
		$\mathbf{s^i} \in \mathbb{Z}_p^r$ representing the local item histogram, where $r$ is the number of items and $s_k^i$ is the count
		for item $k$. The input peers share all elements of their $\mathbf{\mathbf{s^i}}$ among the PPs.
	
	\item \textbf{Summation:} The PPs compute the item counts $[s_k] = \sum_{i=1}^{n}{[s_k^i]}$. Also, the total count~$[S]=\sum_{k=1}^r{[s_k]}$ is computed and reconstructed. 
	
	\item \textbf{Exponentiation:} The PPs compute $[(s_k)^q]$ using square-and-multiply.
	%In our evaluation (see Section~\ref{sec:evaluation}), we set $q=2$.
	
	\item \textbf{Entropy Computation:} The PPs compute the sum $\sigma=\sum_{k}{[(s_k)^q]}$ and reconstruct $\sigma$.
	Finally, at least one PP uses $\sigma$ to (locally) compute the
	Tsallis entropy $H_q(Y) = \frac{1}{q-1} (1- {\sigma}/{S^q})$.
	
	\end{compactenum}
	\end{small}
	\end{boxedminipage}
	\caption{Algorithm for entropy protocol.}
	\label{box:entropy}
\end{figure}

The computation of the entropy of feature distributions has been
successfully applied in network anomaly detection, 
e.g.~\cite{SubspaceMethod05,brauckhoff2009applying,li2006dai,ziviani2007nad}.
Commonly used feature
distributions are, for example, those of IP addresses, port numbers, flow sizes
or host degrees. 
%Particularly delicate is the case of IP addresses or address prefixes, as
%IP addresses often represent individuals.
The Shannon entropy of a feature distribution~$Y$ is $H(Y) =
- \sum_{k}{p_k \cdot \log_{2}(p_k) }$, where~$p_k$ denotes the
probability of an item~$k$. If~$Y$ is a distribution of port numbers,
$p_k$ is the probability of port~$k$ to appear in the traffic
data. The number of flows (or packets) containing item~$k$ is divided
by the overall flow (packet) count to calculate $p_k$. Tsallis
entropy is a generalization of Shannon
entropy that also finds applications in anomaly
detection~\cite{ziviani2007nad,tellenbach2009tsallis}. 
It has been substantially studied with a rich bibliography available in~\cite{tsallis}. 
The 1-parametric Tsallis entropy is defined as:
\begin{equation}
H_q(Y) = \frac{1}{q-1} \Bigl( 1- \sum_{k}{(p_k)^q} \Bigr) . \label{eq:tsallis}
\end{equation}
and has a direct interpretation in terms of moments of order~$q$ of
the distribution. In particular,
the Tsallis entropy is a generalized, non-extensive entropy that, up to a multiplicative constant, 
equals the Shannon entropy for \mbox{$q\rightarrow 1$}. 
%By adapting $q$, it
%allows to specifically look for changes in the distribution of heavy
%hitters or rare events and therefore gives a multi-faceted view on
%distributional changes~\cite{tellenbach2009tsallis}. 
For generality, we select to design an MPC protocol for the Tsallis entropy.

\paragraph{Entropy Protocol}

A straight-forward approach to compute entropy is to first find the
overall feature distribution~$Y$ and then to compute the entropy of
the distribution. In particular, let~$p_k$ be the overall probability
of item~$k$ in the union of the private data and $s_k^i$ the local
count of item~$k$ at input peer~$i$. If~$S$ is the total count of the items,
then $p_k = \frac{1}{S}\sum^n_{i=1}{s_k^i}$.  Thus, to compute the
entropy, the input peers could simply use the addition protocol to add all
the $s_k^i$'s and find the probabilities~$p_k$. Each input peer could then
compute~$H(Y)$ locally. However, the distribution~$Y$ can still be
very sensitive as it contains information for each item, e.g., per 
address prefix. For this reason, we aim at computing~$H(Y)$ without
reconstructing any of the values~$s_k^i$ or~$p_k$. 
Because the rational numbers $p_k$ can not be shared directly over a prime field, we
perform the computation separately on private numerators ($s_k^i$) and the public
overall item count $S$. The entropy protocol achieves this goal as described in Fig.~\ref{box:entropy}.
It is assured that sensitive intermediate results are
not leaked and that input and privacy peers \emph{only} learn the final
entropy value $H_q(Y)$ and the total count $S$. $S$ is not sensitive as it only 
represents the total flow (or packet) count of all input peers together. This can be easily computed 
by applying the addition protocol to volume-based metrics. 
%We therefore assume the input peers are willing to compute $S$.
The complexity of this protocol is $r\log_2 q$ multiplications in $\log_2 q$ rounds.

%The authors of~\cite{tellenbach2009tsallis} compute the
%Traffic Entropy Spectrum (TES) by computing the Tsallis entropy for a
%set of values of $q$. Our current protocol only supports integer values for $q>1$.
%A generalization to rational and negative $q$ values involves handling rational 
%secrets, e.g., by sharing numerator and denominator separately. We leave this for future work.

\subsubsection{Distinct Count}

\begin{figure}[t]
	\begin{boxedminipage}{1\textwidth}
	\begin{small}
	\begin{compactenum}

	\item \textbf{Share Generation:} Each input peer~$i$ shares its negated local counts~$c_k^i = \neg s_k^i$ among the PPs.
	
	%\item \textbf{Verify Shares:} The privacy peers verify that the inputs
	%provided by the peers are in $\{0,1\}$. To verify that a secret
	%input~$c_k^i$ lies in $\{0,1\}$ without disclosing it, they use
	%their shares to jointly compute
	%$c_k^i \cdot (c_k^i - 1)$, which is zero if and only if
	%$c_k^i$ is in $\{0,1\}$. If the input of a peer does not conform,
	%the peer is excluded from the computation.
	
	\item \textbf{Aggregation:} For each item~$k$, the PPs
	compute $[c_k] = [c_k^1] \wedge [c_k^2] \wedge \ldots [c_k^n]$. This can
	be done in $\log_2 n$ rounds. If an item~$k$ is reported 
	by any input peer, then~$c_k$ is $0$.
	
	\item \textbf{Counting:} Finally, the PPs build 
	the sum $[\sigma] = \sum [c_k]$ over all items and reconstruct $\sigma$. 
	The distinct count is then given by $K-\sigma$, where $K$ is the size of the
	item domain.

	\end{compactenum}
	\end{small}
	\end{boxedminipage}
	\caption{Algorithm for distinct count protocol.}
	\label{box:count}
\end{figure}

%The privacy requirements for our distinct count algorithm demand that i)
%the identity of the items seen by a specific peer are private, ii) the
%number of occurrences of an item and the number of peers having seen
%the item remain private and iii) the set of seen items is private.

% Do we need this lengthy discussion of related set operation solutions?
% Adds a lot of references, we operate on Shamir Shares anyway... Compact it!}

%There has been a lot of research on developing algorithms for
%secure set union and secure set operations in general
%(e.g.,~\cite{frikken2007pps,brickell2005ppg,kantarcioglu2004ppd,kissner2005pps}). 
%However, existing solutions are not
%well suited to our needs. The set operations in~\cite{kissner2005pps}
%and~\cite{frikken2007pps} are secure against malicious adversaries
%which is overly complex and therefore less efficient for our
%scenario. 
%The two approaches to secure set union in~\cite{clifton2002tpp} either use general binary circuits or leak intermediate information. 
%The solution from~\cite{brickell2005ppg} was
%proven secure in the semi-honest model but supports only two
%parties. The work in~\cite{kantarcioglu2004ppd} on the other hand,
%leaks superfluous information in order to improve performance.

In this section, we devise a simple distinct count protocol leaking no intermediate
information. 
Let $s_k^i \in \{0,1\}$ be a boolean variable equal to~$1$ if input peer~$i$ sees item~$k$
and $0$ otherwise. We first compute the logical \texttt{OR} of the
boolean variables to find if an item was seen by any input peer or
not. Then, simply summing the number of variables equal to~$1$ gives
the distinct count of the items. 
%According to De Morgan Theorem~\cite{mano2001dd}:
According to De Morgan's Theorem, $a \vee b = \neg(\neg a \wedge \neg b)$.
This means the logical \texttt{OR} can be realized by performing a logical \texttt{AND}
on the negated variables. This is convenient, as the logical \texttt{AND} is
simply the product of two variables. Using this observation, we construct the protocol described in Fig.~\ref{box:count}.
This protocol guarantees that only the distinct count is learned from
the computation; the set of items is \emph{not}
reconstructed. However, if the input peers agree that the item set is not
sensitive it can easily be reconstructed after step 2.
The complexity of this protocol is $(n-1)r$ multiplications in $\log_2 n$ rounds.

\section{Performance Evaluation}
\label{sec:evaluation}

\begin{figure*}[t]
  \centering
  \subfloat[Average round time ($s=30$).] { 
  	\includegraphics[scale=0.5]{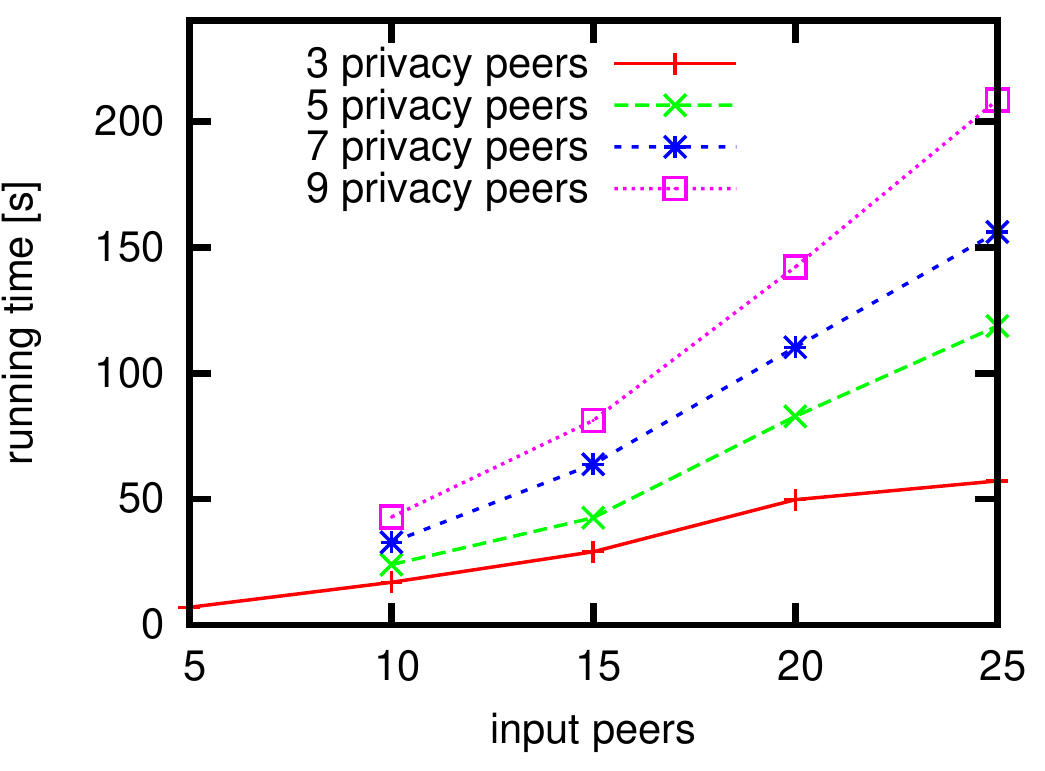} \label{fig:eventeval1}
  }
  \subfloat[Data sent per PP ($s=30$).] {  
  	\includegraphics[scale=0.5]{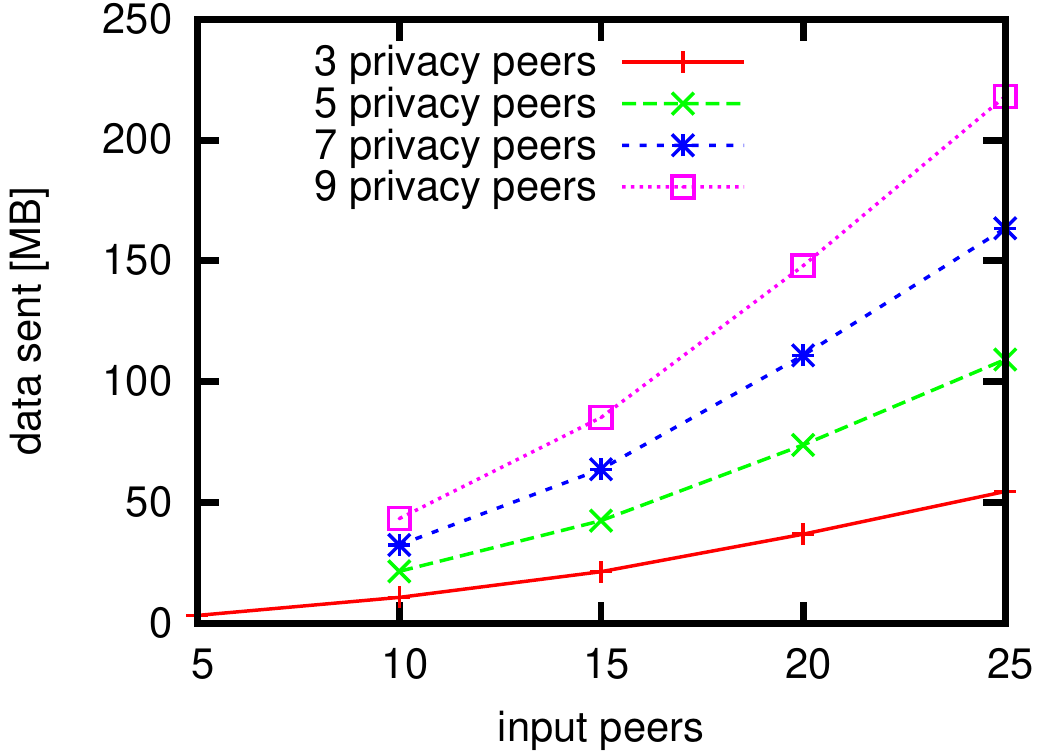} \label{fig:eventeval2}
  }
  \subfloat[Round time vs. $s$ ($n$=10, $m$=3).] {
    \includegraphics[scale=0.5]{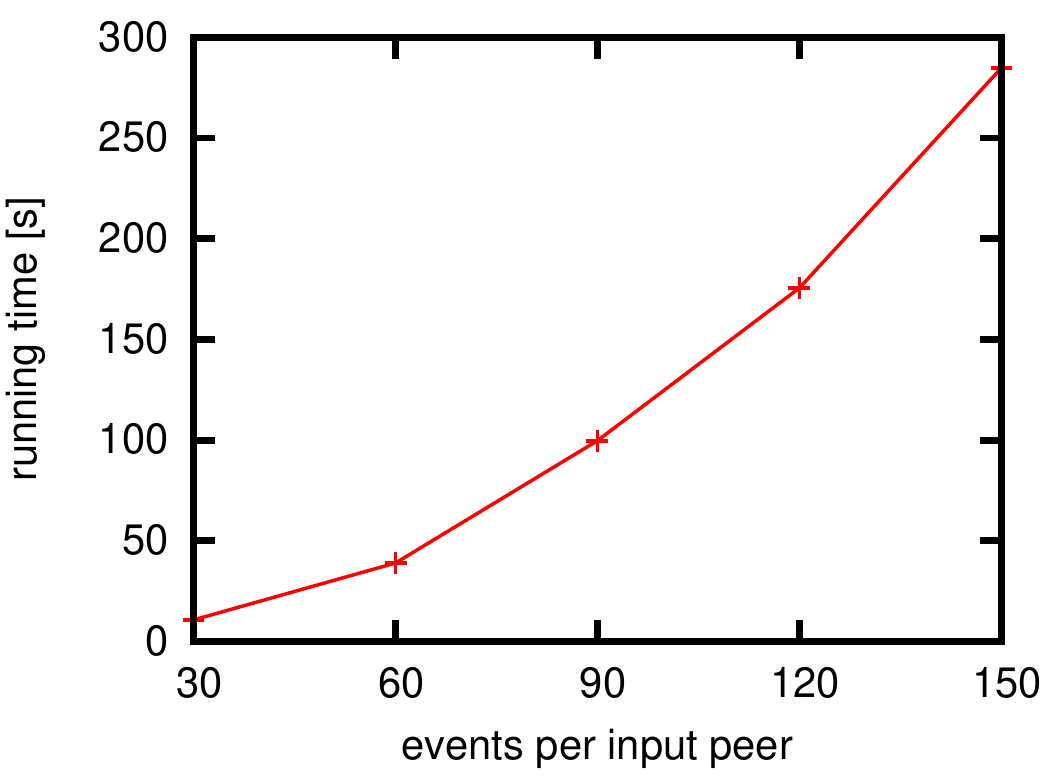} \label{fig:eventevalAlerts} 
  }
	\caption{Round statistics for event correlation with $T_c=n/2$. $s$ is the number of events per input peer. }
	\label{fig:eventeval}
\end{figure*}

In this Section we evaluate the event correlation protocol and the 
protocols for network statistics. After that we explore the impact of
running selected protocols on PlanetLab where hardware, network delay, and bandwidth are very heterogeneous.
This section is concluded with a performance comparison between SEPIA and existing general-purpose MPC frameworks.

We assessed the CPU and network bandwidth requirements of our
protocols, by running different aggregation tasks with real and simulated network data.
For each protocol, we ran several experiments
varying the most important parameters. We varied the number of input peers $n$ between 5 and 25
and the number of privacy peers $m$ between 3 and 9, with $m < n$. The experiments were conducted
on a shared cluster comprised of several public workstations;
each workstation was equipped with a 2x Pentium 4 CPU (3.2\,GHz),
2 GB memory, and 100\,Mb/s network. Each input and privacy peer was run on a \emph{separate} host.
In our plots, each data point reflects the average over 10 time windows. Background load due to user activity could not be totally avoided. 
Section~\ref{sec:eval_planetlab} discusses  
the impact of single slow hosts on the overall running time.

\subsection{Event Correlation} 
\label{sec:eval_event}

For the evaluation of the event correlation protocol, we generated artificial event data. 
It is important to note that our performance metrics do not depend on the
actual values used in the computation, hence artificial data is just as good as real data for these purposes.

\paragraph{Running Time}
Fig.~\ref{fig:eventeval} shows evaluation results for event correlation with $s=30$ events per input peer, each with 24-bit keys for $T_c=n/2$. We ran the protocol including weight and key verification.
Fig.~\ref{fig:eventeval1} shows that the average running time per time window always stays below 3.5\,min and scales quadratically with $n$, as expected. 
%From section~\ref{sec:ec} we know, that with $T_c=n/2$, the number of operations to perform scales quadratically with $n$. 
Investigation of CPU statistics shows that with increasing $n$ also the average CPU load per privacy peer grows. Thus, as long as CPUs are not used to capacity, local parallelization manages to compensate parts of the quadratical increase. With $T_c=n-const$, the running time as well as the number of operations scale linearly with $n$. 
Although the total communication cost grows quadratically with $m$, the running time dependence on $m$ is rather linear, as long as the network is not saturated.  The dependence on the number of events per input peer $s$ is quadratic as expected without optimizations (see Fig.~\ref{fig:eventevalAlerts}).

To study whether privacy peers spend most of their time waiting due to synchronization, we measured the user and system time of their hosts. All the privacy peers were constantly busy with average CPU loads between 120\% and 200\% for the various operations.\footnote{When run on a 32-bit platform, up to twice the CPU load was observed, with similar overall running time. This difference is due to shares being stored in \texttt{long} variables, which are more efficiently processed on 64-bit CPUs.} Communication and computation between PPs is implemented using separate threads to minimize the impact of synchronization on the overall running time. Thus, SEPIA profits from multi-core machines. Average load decreases with increasing need for synchronization from multiplications to $equal$, over $lessThan$ to event correlation. Nevertheless, even with event correlation, processors are very busy and not stalled by the network layer.

%\begin{figure}[t]
  %\centering
  %\includegraphics[angle=270,scale=0.5]{figures/eval/cpuload}
	%\caption{Avg. CPU load for 5 minutes on 64-bit CPUs with 4 cores ($n=3, m=5$).}
	%\label{fig:eventevalCPU} 
%\end{figure}

\paragraph{Bandwidth requirements}
Besides running time, the communication overhead imposed on the network is an important performance measure. 
Since data volume is dominated by privacy peer messages, we show the average bytes sent
\emph{per privacy peer} in one time window in Fig.~\ref{fig:eventeval2}. Similar to
running time, data volume scales roughly quadratically with $n$ and linearly with $m$.
In addition to the transmitted data, each privacy peer receives about the same amount of data from
the other input and private peers. 
If we assume a 5-minute clocking of the event correlation protocol, an average bandwidth between 0.4\,Mbps (for $n=5$, $m=3$) and
13\,Mbps (for $n=25$, $m=9$) is needed per privacy peer.
Assuming a 5-minute interval and sufficient CPU/bandwidth resources, the maximum number of supported input peers before the system stops working in real-time ranges from around 30 up to roughly 100, depending on protocol parameters.

\begin{figure*}[t]
  \centering
  \subfloat[Addition of port histogram.] { 
  	\includegraphics[scale=0.48]{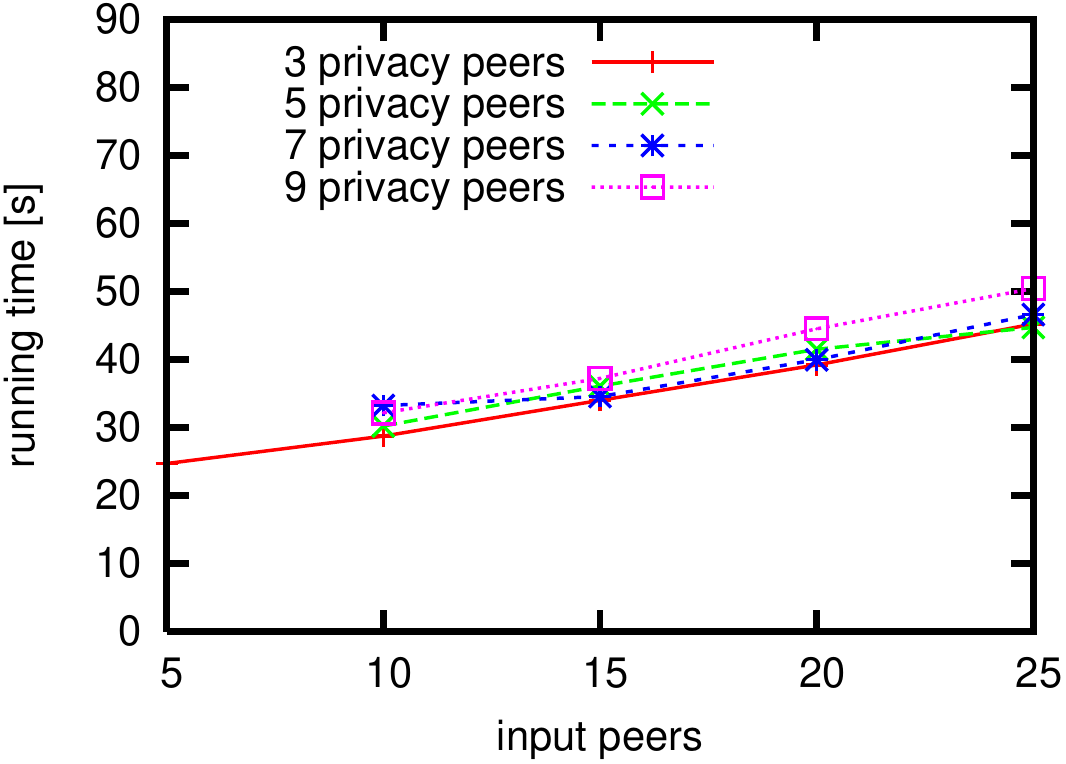}}
	\subfloat[Entropy of port distribution.] { 
		\includegraphics[scale=0.48]{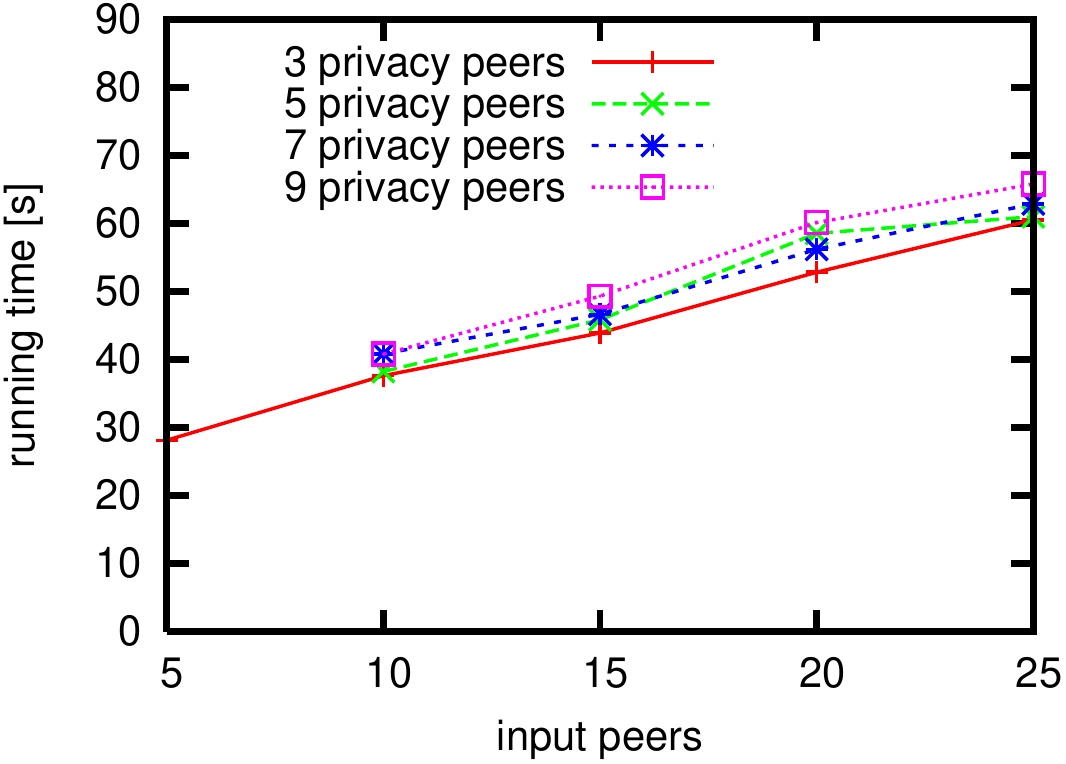} }
	\subfloat[Distinct AS count.] { 
		\includegraphics[scale=0.48]{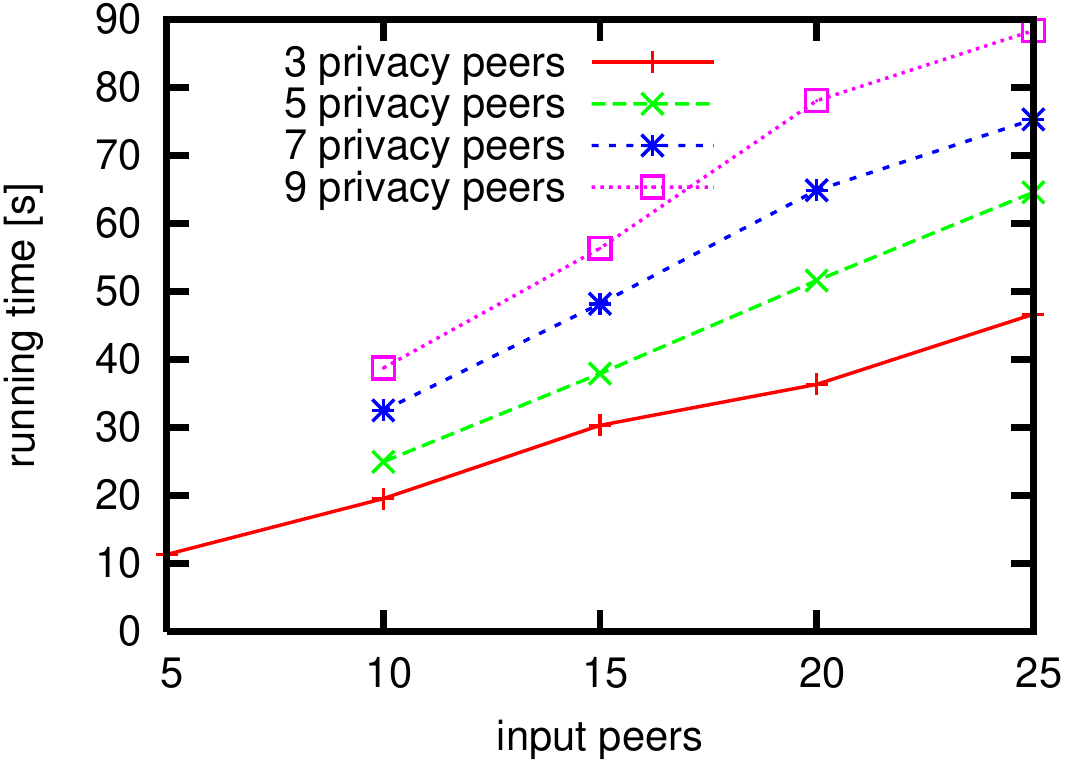} }
	
	\caption{Network statistics: avg. running time per time window versus $n$ and $m$, measured on a department-wide cluster. 
	All tasks	were run with an input set size of 65k items.}
	\label{fig:statistics}
\end{figure*}

\subsection{Network statistics}
\label{sec:eval_stats}

For evaluating the network statistics protocols, we used unsampled NetFlow data captured from 
the five border routers of the Swiss academic and research network (SWITCH), 
a medium-sized backbone operator, connecting
approximately 50 governmental institutions, universities, and research
labs to the Internet. We first extracted traffic
flows belonging to different customers of SWITCH and assigned an independent input peer to each organization's trace. 
For each organization, we then generated SEPIA input files, where each input field contained either the values
of volume metrics to be added or the local histogram of feature
distributions for collaborative entropy (distinct
count) calculation. In this section we focus on the running time and bandwidth requirements only.
%A case study using the same traces for collaborative detection and trouble-shooting of anomalies is presented in Section~\ref{sec:casestudy}.
We performed the following tasks over ten 5-minute windows:

\vspace{1mm}
\begin{compactenum}
\item \textbf{Volume Metrics:} Adding 21 volume metrics containing flow, packet, and
byte counts, both total and separately filtered by protocol (TCP,
UDP, ICMP) and direction (incoming, outgoing). For example,
Fig.~\ref{fig:flows} in Section~\ref{sec:casestudy} plots the total and local number of 
incoming UDP flows of six organizations for an 11-day period. 
%These metrics allow to monitor overall traffic volume and ratios between
%different protocols. Anomalies visible in these metrics are, for instance, outages, 
%alpha flows, scanning, and (D)DoS events. 

\item \textbf{Port Histogram:} Adding the full destination port histogram for
incoming UDP flows. SEPIA input files contained 65,535 fields, 
each indicating the number of flows observed
to the corresponding port. These local histograms were
aggregated using the addition protocol.

\item \textbf{Port Entropy:} Computing the Tsallis entropy of destination ports
for incoming UDP flows. The local SEPIA input files contained the
same information as for histogram aggregation. The Tsallis exponent $q$ was set to $2$.  

\item \textbf{Distinct count of AS numbers:} Aggregating the count of distinct source
AS numbers in incoming UDP traffic. The input files contained
65,535 columns, each denoting if the corresponding source AS
number was observed. 
For this setting, we reduced the field size $p$ to 31 bits because
the expected size of intermediate values is much smaller than for the other tasks.
%An increase or decrease of source AS numbers is a measure of the
%spatial distribution of remote communication points. A drop in the
%number of seen AS numbers could point at potential routing problems,
%e.g., blackholes, whereas a sudden increase is indicative of global
%events or of attacks involving a large number of spoofed source IP
%addresses.

\end{compactenum}
\vspace{1mm}

\paragraph{Running Time}
For task 1, the average running time was below 1.6\,s per time window for
all configurations, even with 25 input and 9 privacy peers. This confirms
that addition-only is very efficient for low volume input data.
Fig.~\ref{fig:statistics} summarizes the running time for tasks 2 to 4. The
plots show on the $y$-axes the average running time per
time window versus the number of input peers on the $x$-axes. 
In all cases, the running time for processing one
time window was below 1.5 minutes.
The running time clearly scales linearly with
$n$. Assuming a 5-minute interval, we can estimate by
extrapolation the maximum number of supported input peers before the system
stops working in real-time. For the conservative case with~9
privacy peers, the supported number of input peers is approximately~140 for histogram addition,
110 for entropy computation, and~75 for distinct count computation.
We observe, that for single round protocols (addition and entropy),  
the number of privacy peers has only little impact on the running time. 
For the distinct count protocol, the running time increases linearly with both $n$ and $m$. 
Note that the shortest running time for distinct count is even lower than for histogram
addition. This is due to the reduced field size ($p$ with 31 bits instead of $62$), which reduces both CPU and network
load.

\paragraph{Bandwidth Requirements}
For all tasks, the data volume sent per privacy peer scales perfectly linear
with $n$ and $m$. Therefore, we only report the maximum volume
with 25 input and 9 privacy peers. For addition of volume metrics, the data volume is 141\,KB and increases
to 4.7\,MB for histogram addition. Entropy computation requires 8.5\,MB and finally the multi-round distinct
count requires 50.5\,MB. For distinct count, to transfer the total of $2 \cdot 50.5=101$\,MB within
5 minutes, an average bandwidth of roughly 2.7\,Mbps is needed per privacy peer.

\subsection{PlanetLab Experiments}
\label{sec:eval_planetlab}

In our evaluation setting hosts have homogeneous CPUs, network bandwidth and low
round trip times (RTT). In practice, however, SEPIA's goal is to aggregate
traffic from remote network domains, possibly resulting in a much more heterogeneous setting.
For instance, \emph{high delay} and \emph{low bandwidth} directly affect the waiting time for
messages. Once data
has arrived, the \emph{CPU model and clock rate} determine how fast the data is processed
and can be distributed for the next round.

%Hence the impact of high delays grows
%with the number of messages sent. Generally, the number of generated
%messages per window is proportional to the total data volume
%generated and scales linearly with the number of peers and privacy peers.

Recall from Section~\ref{sec:protocols} that each operation and protocol in SEPIA
is designed in rounds.
Communication and computation during each round run in parallel. But before the next round can start, the privacy peers
have to synchronize intermediate results and
therefore wait for the slowest privacy peer to finish. The overall running time
of SEPIA protocols is thus affected by the slowest CPU, the highest
delay, and the lowest bandwidth rather than by the average performance of hosts and links.
Therefore we were interested to see whether the performance of our protocols
breaks down if we take it out of the homogeneous LAN setting.
Hence, we ran SEPIA on PlanetLab~\cite{planetlab} and repeated task 4 (distinct AS count) with
10 input and 5 privacy peers on globally distributed PlanetLab nodes.
%in South and North America, Europe, and Asia. 
Table~\ref{tab:planetlab} compares the LAN setup with two PlanetLab setups A and B.

\begin{table}[t]
	\centering
	\begin{minipage}{0.49\textwidth}
		\vspace{0pt}
		\centering
		\begin{small}
		\setlength{\tabcolsep}{3pt}
		\begin{tabular}{p{1.9cm}|ccc}
										 & LAN  	 					& PlanetLab A  					& PlanetLab B \\
			\hline
			Max. RTT 			 & 1\,ms  					&  320\,ms    					& 320\,ms    \\
			Bandwidth		   & 100\,Mb/s  	  	&  $\geq 100$\,Kb/s 	  & $\geq 100$\,Kb/s  \\
			Slowest CPU 	 & 2 cores        	&  2 cores            	& 1 core  	  \\
										 & 3.2\,GHz         &  2.4\,GHz						  & 1.8\,GHz  \\
 		  Running time 	 & 25.0\,s  				&  36.8\,s    					& 110.4\,s   \\
		\end{tabular}
		\end{small}
		\caption{Comparison of LAN and PlanetLab settings.}
		\label{tab:planetlab}
	\end{minipage}
	\hfill
	\begin{minipage}{0.49\textwidth}
		\centering
		\begin{small}
		\setlength{\tabcolsep}{3pt}
		\begin{tabular}{p{1.7cm}|ccc}
			Framework       &  SEPIA         &  VIFF  				& FairplayMP \\
			\hline
			Technique			  &  Shamir sh. & Shamir sh.  & Bool. circuits  \\	      
			Platform				&  Java       & Python      &   Java          \\       
			Multipl./s      &  82,730       &    326    	&    1.6     \\
			Equals/s  	&   2,070 	&     2.4   	&    2.3     \\
			LessThans/s    	&     86   &     2.4       &    2.3     \\
		\end{tabular}
		\end{small}
		\caption{Comparison of frameworks performance in operations per second with $m=5$. }
		\label{tab:frameworks}
	\end{minipage}
\end{table}

RTT was much higher and average bandwidth much lower on PlanetLab. 
The only difference between PlanetLab A and B was the choice of some nodes with slower CPUs. Despite the very heterogeneous
and globally distributed setting, the distinct count protocol performed well, at least in PlanetLab A.
Most important, it still met our near real-time requirements.
From PlanetLab A to B, running time went up by a factor of 3. However, this can largely be explained by the slower CPUs.
The distinct count protocol consists of parallel multiplications, which make efficient use of the CPU and local addition, which is solely CPU-bound. 
Let us assume, for simplicity, that clock rates translate directly into MIPS. Then, computational power in PlanetLab B is roughly 2.7 times lower than in PlanetLab A.  
Of course, the more rounds a protocol has, the bigger is the impact of RTT. But in each round, the network delay is
only a constant offset and can be amortized over the number of parallel operations performed per round. For 
many operations, CPU and bandwidth are the real bottlenecks.

While aggregation in a heterogeneous environment is possible, SEPIA \emph{privacy peers} should ideally be deployed on dedicated hardware, to reduce background load, and with similar CPU equipment, so that no single host slows down the entire process.

\subsection{Comparison with General-Purpose Frameworks}
\label{sec:eval_frameworks}

% Sharemind (3 PPs only!)
% ----------------
% multi: ~163,000/s 
% lessThan: ~360/s 
%
%
% Measured on TARDIS-C with benchmark protocol 
% ----------------------------------------------------
%
% Privacy Peers:	5 
% mpc.benchmark.shamirsharesfieldorder=1401085391
% mult:  	
% equals:   1,570 (with MATLAB) 				[50,000 per round]
% lessthans: 79 / 403 (without random) 	[ 3,000 per round]
%

% Measured on CLUSTER with benchmark protocol (BasicMetrics running on all nodes)
% -----------------------------------------------------
%
% Privacy Peers:	5 
% mpc.benchmark.shamirsharesfieldorder=1401085391
% mult:  	82,735
% equals:  2,070
% lessthans: 86 / 381 (without random)
%
% Privacy Peers:	3
% mpc.benchmark.shamirsharesfieldorder=1401085391
% mult:  	142,795
% equals:   3,377
% lessthans: 143 / 614 (without random)

% -> slower than with WSIP on tardis-c? How can this be?

In this section we compare the performance of basic SEPIA operations to those of general-purpose frameworks such as FairplayMP~\cite{fairplayMP} and VIFF v0.7.1~\cite{damgard2009asynchronous}. Besides performance, one aspect to consider is, of course, usability. Whereas the SEPIA library currently only provides an API to developers, FairplayMP allows to write protocols in a high-level language called SFDL and VIFF integrates nicely into the Python language. Furthermore, VIFF implements asynchronous protocols and provides plenty of additional modules, including security against malicious adversaries and for MPC based on homomorphic cryptosystems. 

Tests were run on 2x Dual Core AMD Opteron 275 machines with 1Gb/s LAN connections. For all frameworks, the semi-honest model, 5 computation nodes, and $32$ bit input numbers were used. Table~\ref{tab:frameworks} shows the average number of parallel operations per second for each framework.
SEPIA clearly outperforms VIFF and FairplayMP for all operations and is thus much better suited when performance of parallel operations is of main importance.
As an example, a run of event correlation taking 3 minutes with SEPIA would take roughly 2 days with VIFF. This extends the range of \emph{practically} runnable MPC protocols significantly. Notably, SEPIA's $equal$ operation is 24 times faster than its $lessThan$, which requires 24 times more multiplications, but at the same time also twice the number of rounds. This confirms that with many parallel operations, the number of multiplications becomes the dominating factor.
Approximately $3/4$ of the time spent for $lessThan$ is used for generating sharings of random numbers used in the protocol. These random sharings are independent from input data and could be generated prior to the actual computation, allowing to perform 380 $lessThan$s per second in the same setting.

%The results confirm that the garbled boolean circuit approach used by FairplayMP is better suited to perform comparison than arithmetic operations, such as multiplications. For arithmetic (Shamir) sharing it is exactly the other way round. VIFF outperforms FairplayMP for multiplications and matches it for comparison operations.

Even for multiplications, SEPIA is faster than VIFF, although both rely on the same scheme. We assume this can largely be attributed to the completely asynchronous protocols implemented in VIFF. Whereas asynchronous protocols are very efficient for dealing with malicious adversaries, they make it impossible
to reduce network overhead by exchanging intermediate results of all parallel operations at once in a single big message. Also, there seems to be a bottleneck in parallelizing large numbers of operations. In fact, when benchmarking VIFF, we noticed that after some point, adding more parallel operations significantly slowed down the average running time per operation. 

Sharemind~\cite{Sharemind} is another interesting MPC framework using \emph{additive} secret sharing to implement multiplications and greater-or-equal (GTE) comparison. The authors implement it in C++ to maximize performance. However, the use of additive secret sharing makes the implementations of basic operations dependent on the number of computation nodes used. For this reason, Sharemind is currently restricted to $3$ computation nodes only. Regarding performance, however, Sharemind is comparable to SEPIA. According to~\cite{Sharemind}, Sharemind performs up to 160,000 multiplications and around 330 GTE operations per second, with 3 computation nodes. With 3 PPs, SEPIA performs around 145,000 multiplications and 145 $lessThan$s per second (615 with pre-generated randomness). Sharemind does not directly implement $equal$, but it could be implemented using 2 invocations of GTE, leading to $\approx 115$ operations/s. SEPIA's $equal$ is clearly faster with up to $3,400$ invocations/s.
SEPIA demonstrates that operations based on Shamir shares are not necessarily slower than operations in the additive sharing scheme. The key to performance is rather an implementation, which is optimized for a large number of parallel operations. Thus, SEPIA combines speed with the flexibility of Shamir shares, which support any number of computation nodes and are to a certain degree robust against node failures. 

\section{Design and Implementation}
\label{sec:design}
\lstset{language=Java,numbers=none,frame=single,basicstyle=\ttfamily\tiny,breaklines,tabsize=2,showstringspaces=false}

%\begin{figure}[t]
	%\centering
	%\includegraphics[clip,scale=0.68]{figures/architecture}
	%\caption{The SEPIA library and command line tools.}
	%\label{fig:architecture}
%\end{figure}

% architecture

The foundation of the SEPIA library is an implementation of the basic operations, such as multiplications and optimized comparisons 
(see Section~\ref{sec:comparisons}), along
with a communication layer providing a peer-to-peer infrastructure over secure channels, realized by
SSL connections.
In order to limit the impact of varying communication latencies and response
times, each connection, along with the corresponding computation and communication tasks, is
handled by a separate thread. This also implies that SEPIA protocols benefit from multi-core systems for computation-intensive tasks.
In order to reduce synchronization overhead, intermediate results of parallel operations sent to the same destination are collected and transfered in one big message instead of many small messages.
On top of the basic layers, the protocols from Section~\ref{sec:protocols} are implemented as standalone command-line (CLI) tools. 
The CLI tools expect a local configuration file containing privacy peer addresses, paths to a folder with input data and a Java keystore, as well as protocol-dependent parameters. 
The tools write a detailed log of the ongoing computation and output files with aggregate results for each time window.
The keystore holds certificates of trusted input and privacy peers to establish SSL connections.
It is possible to delay the start of a computation until a (configurable) minimum number of input and privacy peers are online. 
This gives the input peers the ability to define an acceptable level of privacy by 
only participating in the computation if a certain minimum number of other input/privacy peers also participate.

% implementation / API

SEPIA is written in Java to provide platform independence. 
The source code of the basic library and the four CLI tools is available under the LGPL license. 
There one can also find pre-configured examples for the CLI tools and a user manual. The user manual describes usage and configuration of the CLI tools and includes a step-by-step
tutorial on how to use the library API to develop new protocols.
In the library API, all operations and subprotocols implement a common interface 
\texttt{IOperation} and are easily composable. The class \texttt{Protocol\-Primitives} 
allows to schedule operations and takes care of performing them in parallel, keeping track of operation states. A base class for privacy peers 
implements the \texttt{doOperations()} method, which runs all the necessary computation rounds
and synchronizes data between privacy peers in each round. Fig.~\ref{fig:code} shows example code for three input peers that want to privately compare their secrets.
First, each input peer generates shares of its secret. The shares are then sent to the PPs. The PPs first schedule and execute \emph{lessThan} comparisons for all combinations of input secrets. In a second step, they run the reconstruction operations and output the results.

% usage
%Each input and privacy peer maintains a list of trusted privacy peers and
%their certificates in a Java Keystore (JKS). Addresses and ports of privacy peers are stored in the
%local configuration file and new privacy peers have to be added
%manually by the administrator. 
%For an input peer, it is sufficient to know the address of at least one privacy peer; once contacted, the privacy peer
%will forward a list of all participating privacy peers to the peer.

\begin{figure}
\begin{tabular}{ll}
	\begin{minipage}[t]{0.39\textwidth}
	Input peer 1 (other IPs do the same):
	\begin{lstlisting}
ShamirSharing sharing = new ShamirSharing();
sharing.setFieldPrime(1401085391); // 31 bit
sharing.setNrOfPrivacyPeers(nrOfPrivacyPeers); 
sharing.init();

// Secret1: only a single value
long[] secrets = new long[]{1234567}; 
long[][] shares = sharing.generateShares(secrets);

// Send shares to each privacy peer
for(int i=0; i<nrOfPrivacyPeers; i++) {
	connection[i].sendMessage(shares[i]);
}
	\end{lstlisting}
	\end{minipage}
&	
	\begin{minipage}[t]{0.57\textwidth}
	Privacy peer 1 (other PPs do the same):
	\begin{lstlisting}
... // receive all the shares from input peers
ProtocolPrimitives primitives = new ProtocolPrimitives(fieldPrime, ...);

// Schedule comparisons of all the input peer's secrets
int id1=1, id2=2, id3=3; // consecutive operation IDs
primitives.lessThan(id1, new long[]{shareOfSecret1, shareOfSecret2});
primitives.lessThan(id2, new long[]{shareOfSecret2, shareOfSecret3});
primitives.lessThan(id3, new long[]{shareOfSecret1, shareOfSecret3});
doOperations(); // Process operations and sychronize intermediate results

// Get shares of the comparison results
long shareOfLessThan12 = primitives.getResult(id1);
long shareOfLessThan23 = primitives.getResult(id2);
long shareOfLessThan13 = primitives.getResult(id3);
  
// Schedule and perform reconstruction of comparisons
primitives.reconstruct(id1, new long[]{shareOfLessThan12});
primitives.reconstruct(id2, new long[]{shareOfLessThan23});
primitives.reconstruct(id3, new long[]{shareOfLessThan13});
doOperations();

boolean secret1_lessThan_secret2 = (primitives.getResult(id1)==1);
boolean secret2_lessThan_secret3 = (primitives.getResult(id2)==1);
boolean secret1_lessThan_secret3 = (primitives.getResult(id3)==1);
	\end{lstlisting}
	\end{minipage}
\end{tabular} 
\caption{Example code using the SEPIA library. On the left, input peers provide a secret, e.g., three millionaires sharing their amount of wealth. The privacy peers (right side) then privately compare these values, e.g., to find who is the richest, and reconstruct the comparison results without learning the secrets.}%
\label{fig:code}%
\end{figure}

\paragraph{Future Work}
Note that with Shamir shares, computation can continue and reconstruction of results is assured as long as $t+1$ PPs are online and responsive. This can be used directly to extend SEPIA protocols with robustness against node failures. Also, weak nodes slowing down the entire computation could be excluded from the computation. We leave this as a future extension.

The protocols support any number of input and privacy
peers. Also, the item set sizes/events per input peer are configurable and thus only
limited by the available CPU/bandwidth resources. 
However, running the network statistics protocols (e.g., distinct count) on very large distributions, such as the global IP address range, requires to use sketches as proposed in~\cite{roughan2006sdd} or binning (e.g., use address prefixes instead of addresses).
As part of future work, we plan to investigate the applicability of polynomial set representation to our statistics protocols, to reduce the linear dependency on the input set domain. Polynomial set representation, introduced by Freedman \emph{et al.}~\cite{freedman2004pm} and extended by Kissner 
\emph{et al.}~\cite{kissner2005pps}, represents set elements as roots of a polynomial and enables set operations that scale only logarithmically with input domain size.
However, these solutions use homomorphic public-key cryptosystems, which come with significant overhead for basic operations. Furthermore, they do not trivially allow to separate roles into input and privacy peers, as each input provider is required to 
perform certain non-delegable processing steps on its own data.

\section{Applications}
\label{sec:applications}

% TODO: Put more emphasis on security. Try to cite some Usenix Security papers.

%Providing organizations with incentives to share data in a privacy-preserving
%manner is a challenging problem, for it is typically safer for an organization
%to simply avoid sharing. However, as privacy preserving technologies become
%more mature and well-known, this is likely to change and to open new directions
%for multi-domain privacy-preserving analysis applications.

We envision four distinct aggregation scenarios using SEPIA. The first
scenario is aggregating information coming from multiple domains of
one large (international) organization. This aggregation is presently
not always possible due to privacy concerns and heterogeneous jurisdiction. The
second scenario is analyzing private data owned by three or more
independent organizations with a mutual benefit in collaborating. Five
local ISPs, for example, might collaborate to detect attacks. A third
scenario provides access to researchers for evaluating and validating
traffic analysis or event correlation prototypes over multi-domain
network data. For example, national research, educational, and
university networks could provide SEPIA input and/or privacy peers that allow
analyzing local data according to submitted MPC scripts. Finally,
one last scenario is the privacy-preserving analysis of end-user data,
i.e., end-user workstations can use SEPIA to collaboratively analyze
and cross-correlate local data. 

\subsection{Application Taxonomy}
Based on these scenarios, we see three different classes of possible SEPIA applications.

\paragraph{Network Security}

Over the last years, considerable research efforts have focused on
distributed data aggregation and correlation systems for the
identification and mitigation of coordinated wide-scale attacks. In
particular, aggregation enables the (early) detection and
characterization of attacks spanning multiple domains using data from
IDSes, firewalls, and other possible
sources~\cite{ATLAS,DShield,lincoln2004alerts,yegneswaran2004gid}.
Recent studies~\cite{katti2005cac} show that coordinated wide-scale
attacks are prevalent: 20\% of the studied malicious addresses and
40\% of the IDS alerts accounted for coordinated wide-scale attacks.
Furthermore, strongly correlated groups profiting most from
collaboration have less than 10 members and are stable over time,
which is well suited for SEPIA protocols.

In order to counter such attacks, Yegneswaran \emph{et
al.}~\cite{yegneswaran2004gid} presented DOMINO, a distributed IDS
that enables collaboration among nodes. They evaluated the performance
of DOMINO with a large set of IDS logs from over 1600 providers.
Their analysis demonstrates the significant benefit that is obtained
by correlating the data from several distributed intrusion data
sources. The major issue faced by such correlation systems is the lack
of data privacy. In their work, Porras \emph{et al.}~survey existing
defense mechanisms and propose several remaining research challenges
\cite{porras2006lcs}.  Specifically, they point out the need for
efficient privacy-preserving data mining algorithms that enable
traffic classification, signature extraction, and propagation
analysis.

\paragraph{Profiling and Performance Analysis}

A second category of applications relates to traffic profiling and
performance measurements. A global profile of traffic trends helps
organizations to cross-correlate local traffic trends and identify
changes. In~\cite{sekar2004tfi} the authors estimate that 50 of the
top-degree ASes together cover approximately 90\% of global AS-paths.
Hence, if large ASes collaborate, the computation of global Internet
statistics is within reach. One possible statistic is the total
traffic volume across a large number of networks. This statistic, for
example, could have helped~\cite{roughan2006sdd} in the dot-com bubble
in the late nineties, since the traffic growth rate was overestimated
by a factor of 10, easing the flow of venture capital to Internet
start-ups. In addition, performance-related applications can benefit
from an ``on average'' view across multiple domains. Data from
multiple domains can also help to locate with higher confidence a
remote outage, and to trigger proper detour mechanisms. A number of
additional MPC applications related to performance monitoring are
discussed in~\cite{roughan2006ppp}.

\paragraph{Research Validation}

Many studies are obliged to avoid rigorous validation or have to
re-use a small number of old traffic
traces~\cite{claffy2006con,slagell2005scn}.  This situation clearly
undermines the reliability of the derived results. In this context,
SEPIA can be used to establish a privacy-preserving infrastructure for
research validation purposes. For example, researchers could provide
MPC scripts to SEPIA nodes running at universities and research
institutes.

\subsection{Case Study: Detecting the Skype Outage}
\label{sec:casestudy}

%% SEPIA finds many applications in the fields of network measurements,
%% management, and security. But how big is the potential of
%% collaboration in these fields? From~\cite{katti2005cac} we know that
%% there is great potential in correlating IDS alerts as 40\% of the
%% studied alerts were triggered by attack sources visible on more than
%% one site.  SEPIA greatly enables the study of network event and
%% anomaly correlation between multiple sites and the aggregated view,
%% but performing an in-depth analysis of this question is beyond the
%% scope of this paper.

The Skype outage in August 2007 started from a
Windows update triggering a large number of system restarts. In
response, Skype nodes scanned cached host-lists to find
supernodes causing a huge distributed scanning event lasting two
days~\cite{rossi2009understanding}. We used NetFlow traces of the actual up- and downstream traffic 
of the $17$ biggest customers of the SWITCH network. The traces span~11 days from the 11th to 22nd and include
the Skype outage (on the 16th/17th) along with other smaller
anomalies. We ran SEPIA's total count, distinct count, and entropy 
protocols on these traces and investigated how the organizations can
benefit by correlating their local view with the aggregate view.

%% @martin: where there any ``catastrophic'' consequences of the outage
%% for the organizations? like services going down and money lost? Could
%% you check if any study analyzed such aspects of the outage?

\begin{figure}[t]
	\centering 
	\begin{minipage}{0.6\textwidth}
		\includegraphics[clip,width=1\textwidth]{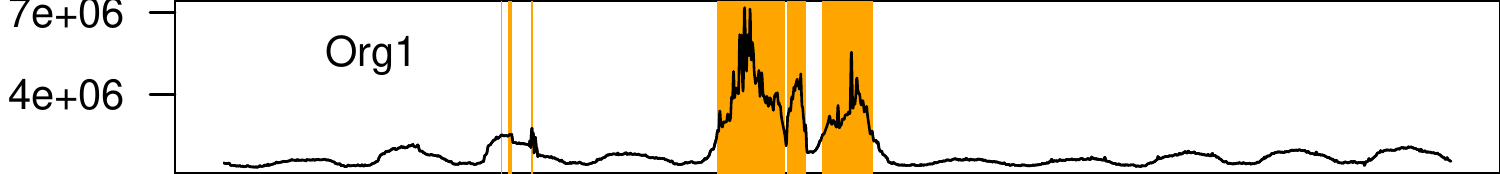} \vspace{-1pt}
		\includegraphics[clip,width=1\textwidth]{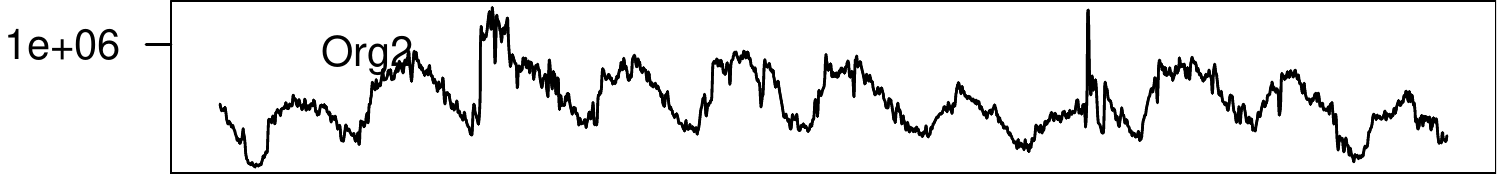} \vspace{-1pt}
		\includegraphics[clip,width=1\textwidth]{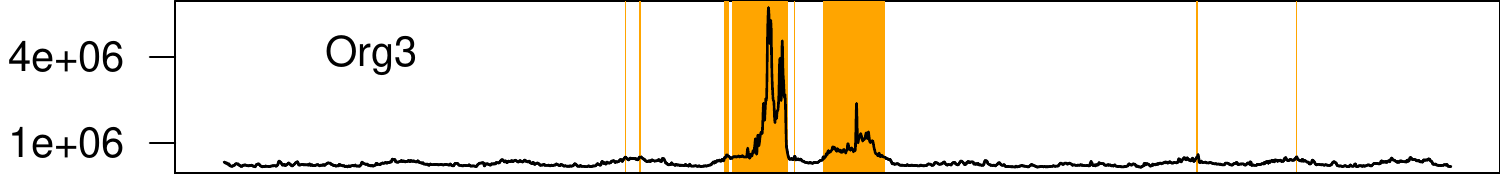} \vspace{-1pt}
		\includegraphics[clip,width=1\textwidth]{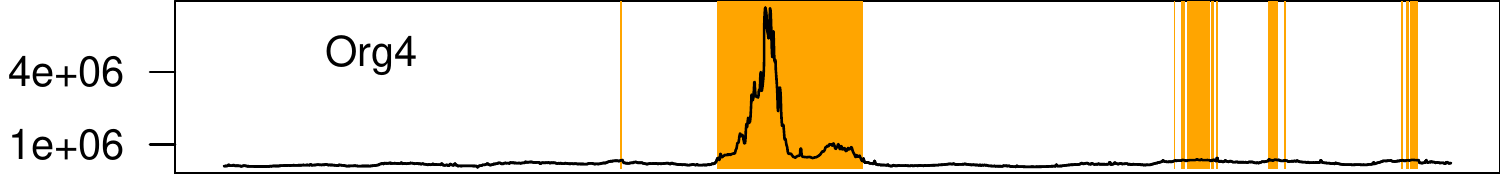} \vspace{-1pt}
		\includegraphics[clip,width=1\textwidth]{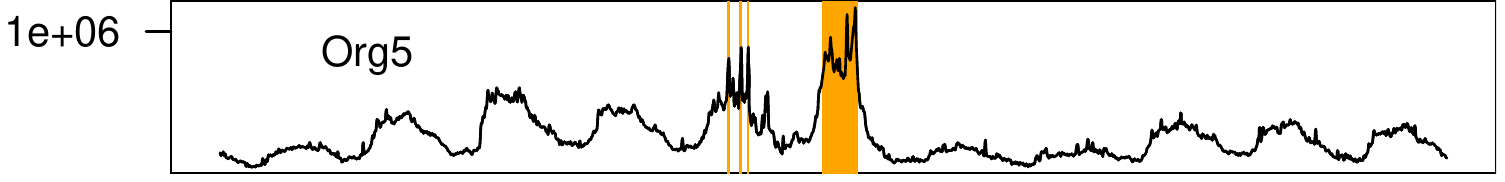} \vspace{-1pt}
		\includegraphics[clip,width=1\textwidth]{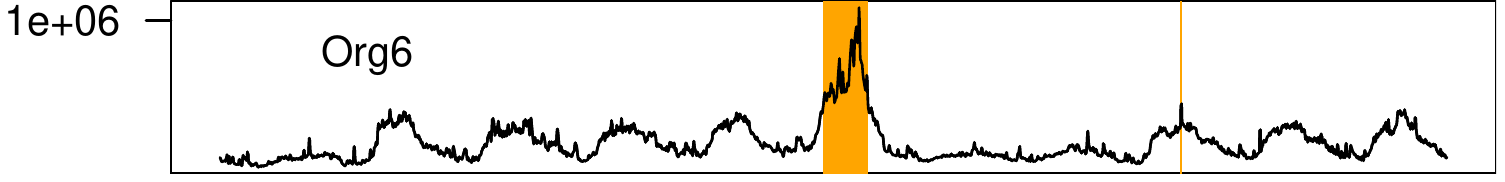} \vspace{-1pt}
		\includegraphics[clip,width=1\textwidth]{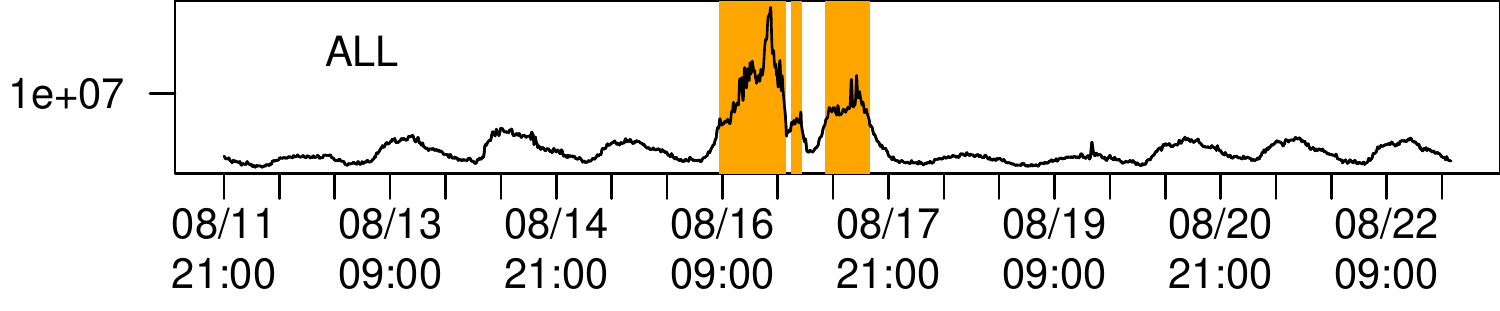}
	\end{minipage}
	\caption{Flow count in 5' windows with anomalies for the biggest organizations and aggregate view (ALL). Note that
		each organization only sees its local and the aggregate traffic.}
	\label{fig:flows}
\end{figure}

We first computed per-organization and aggregate timeseries of the UDP
flow count metric and applied a simple detector to identify
anomalies. For each timeseries, we used the first~4 days to learn its
mean~$\mu$ and standard deviation~$\sigma$, defined the normal region
to be within~$\mu \pm 3\sigma$, and detected anomalous time intervals. In
Fig.~\ref{fig:flows} we illustrate the local timeseries for the six
largest organizations and the aggregate timeseries. We have ranked
organizations based on their decreasing average number of daily flows
and use their rank to identify them. In the figure, we also mark the
detected anomalous intervals. Observe that in addition to the Skype
outage, some organizations detect other smaller anomalies that took
place during the 11-day period.

%\begin{figure}[t]
%	\centering
%	\includegraphics[clip,angle=270,scale=0.48]{figures/incomingFlowCountUDP}
%	\caption{Incoming UDP Flow count per organization and	aggregated }
%	\label{fig:flows}
%\end{figure}

\begin{figure}[t]
	\centering
	\includegraphics[clip,scale=0.45]{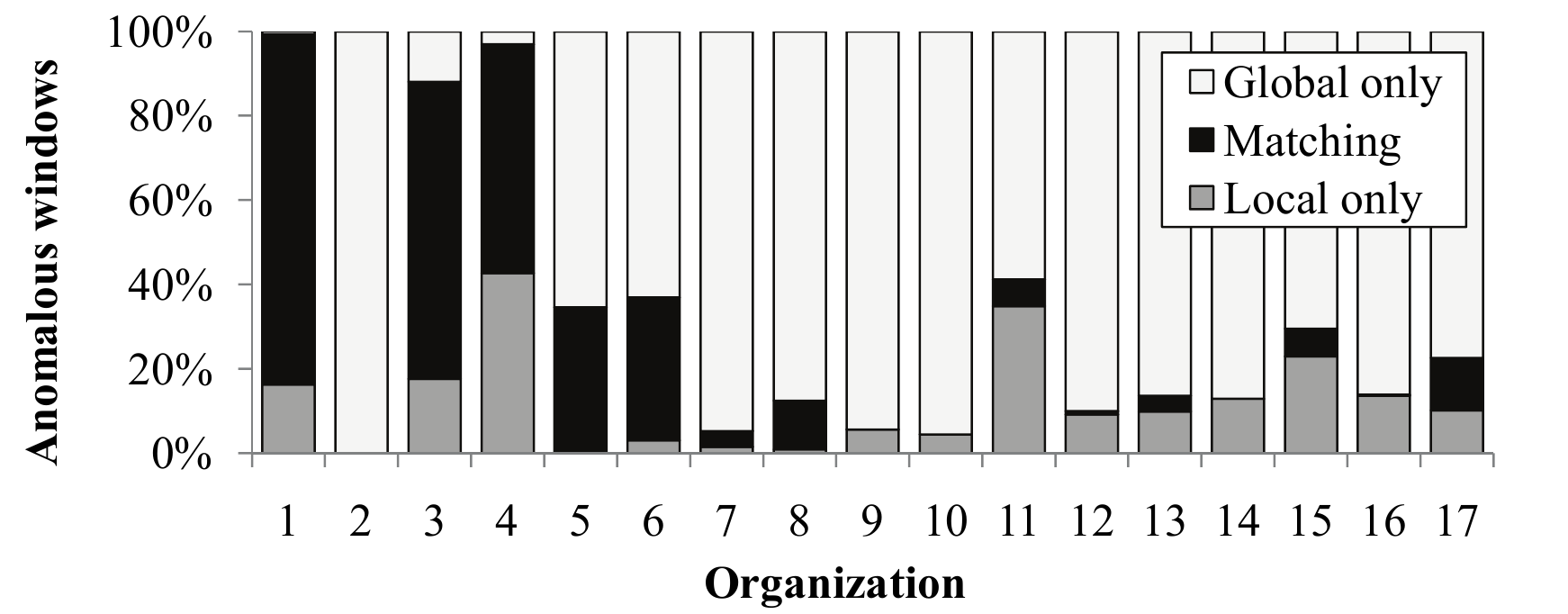}
	\caption{Correlation of local and global anomalies for organizations ordered by size (1=biggest).}
	\label{fig:anomalyCorrelation} 
\end{figure}

%\begin{figure}[t]
%	\centering
%	\includegraphics[clip,scale=0.55]{figures/anomalySizes}
%	\caption{Relative anomaly sizes.}
%	\label{fig:anomalySizes} 
%\end{figure}

%% \begin{figure*}[t]
%% 	\centering
%% 	\subfloat[Tsallis entropy ($q=2$) of destination ports.] {
%% 		\label{fig:entropy}	\includegraphics[clip,angle=270,scale=0.48]{figures/entropyDstPortsIncomingUDP} 
%% 	}
%% 	\subfloat[Distinct count of source AS numbers] {
%% 		\label{fig:counts} \includegraphics[clip,angle=270,scale=0.48]{figures/incomingUniqueCountAS} 
%% 	}

%% 	\caption{Port entropy and AS count in incoming UDP traffic,
%% 	per organization and aggregated.}
%% \end{figure*}

\paragraph{Anomaly Correlation}

Using the aggregate view, an organization can find if a local anomaly
is the result of a global event that may affect multiple
organizations. Knowing the global or local nature of an anomaly is
important for steering further troubleshooting steps. Therefore, we
first investigate how the local and global anomalous intervals
correlate. For each organization, we compared the local and aggregate
anomalous intervals and measured the total time an anomaly was
present: 1) only in the local view, 2) only in the aggregate view, and
3) both in the local and aggregate views, i.e., the {\it matching
anomalous intervals}. Fig.~\ref{fig:anomalyCorrelation} illustrates
the corresponding time fractions. We
observe a rather small fraction, i.e., on average 14.1\%, of
local-only anomalies. Such anomalies lead administrators to
search for local targeted attacks, misconfigured or compromised
internal systems, misbehaving users, etc.  In addition, we observe an
average of 20.3\% matching anomalous windows. Knowing an anomaly is
both local and global steers an affected organization to search
for possible problems in popular services, in widely-used software,
like Skype in this case, or in the upstream providers.
A large fraction (65.6\%) of
anomalous windows is only visible in the global view.
In addition, we observe significant variability in the patterns of
different organizations. In general, larger organizations tend to have
a larger fraction of matching anomalies, as they contribute more to
the aggregate view. While some organizations are highly correlated
with the global view, e.g., organization 3 that notably contributes
only 7.4\% of the total traffic; other organizations are barely
correlated, e.g., organizations 9 and 12; and organization 2 has no
local anomalies at all.

%Fig.~\ref{fig:anomalySizes} plots the relative anomaly size
%of the organizations during the Skype outage. 

\paragraph{Anomaly Troubleshooting}
We define {\it relative anomaly size} to be the ratio of the detection
metric value during an anomalous interval over the detection
threshold.  Organizations~3 and~4 had relative anomaly sizes~11.7
and~18.8, which is significantly higher than the average of~2.6. Using the
average statistic, organizations can compare the relative impact of an
attack. Organization~2, for instance, had anomaly size~0 and 
concludes that there was a large anomaly taking place but they were not
affected.  Most of the organizations conclude that they were
indeed affected, but less than average.  Organizations~3 and~4,
however, have to spend thoughts on why the anomaly was so
disproportionately strong in their networks.

An investigation of the full port distribution and its entropy (plots
omitted due to space limitations) shows that affected organizations
see a sudden increase in scanning activity on specific high port
numbers. Connections originate mainly from ports~80 and~443, i.e.,
the fallback ports of Skype, and a series of high port numbers
indicating an anomaly related to Skype. For organizations~3 and~4,
some of the scanned high ports are extremely prevalent, i.e., a single
destination port accounts for 93\% of all flows at the peak rate.
Moreover, most of the anomalous flows within organizations~3 and~4 are
targeted at a single IP address and originate from thousands of
distinct source addresses connecting repeatedly up to 13 times per
minute. These patterns indicate that the two organizations host
popular supernodes, attracting a lot of traffic to specific
ports. Other organizations mainly host client nodes and see uniform
scanning, while organization~2 has banned Skype completely. Based on
this analysis, organizations can take appropriate measures to
mitigate the impact of the 2-day outage, like notifying users or
blocking specific port numbers.

\paragraph{Early-Warning}
Finally, we investigate whether the aggregate view can be useful for
building an early-warning system for global or large-scale anomalies.
The Skype anomaly did not start concurrently in all locations, which
is often the case with global anomalies, since the Windows update
policy and reboot times were different across organizations. We
measured the lag between the time the Skype anomaly was first observed
in the aggregate and local view of each organization. In
Table~\ref{tab:earlywarning} we list the organizations that had
considerable lag, i.e., above an hour. Notably, one of the most
affected organizations (6) could have learned the anomaly almost one day ahead.
%, which would provide
%sufficient time to take appropriate countermeasures. 
However, as shown in Fig.~\ref{fig:anomalyCorrelation}, for
organization 2 this would have been a false positive alarm. To profit
most from such an early warning system in practice, the aggregate view
should be annotated with additional information, like the number of
organizations or the type of services affected from the same
anomaly. In this context, our event correlation protocol is
useful to find if the same anomaly signatures are observed in the
participating networks. Anomaly signatures can be extracted
automatically using actively researched
techniques~\cite{Brauckhoff2009a,Ranjan2007}.

\begin{table}[ht]

\centering
	\begin{tabular}{l|ccccccc}
		 Org \#     & 3    &  5    & 6     &   7   &  13   &  17   \\
		 \hline
		 lag [hours]    & 1.2  &  2.7  & 23.4  &  15.5 &  4.8  &  3.6   
	\end{tabular}

\caption{Organizations profiting from an early anomaly
	warning by aggregation.}

\label{tab:earlywarning}
\end{table}

%% involve groups of organizations offering similar
%% services and therefore being exposed to similar threats (see
%% also~\cite{katti2005cac}).

%% Consulting other statistics, such as the port entropy
%% (Fig.~\ref{fig:entropy}) or AS number count (Fig.~\ref{fig:counts})
%% gives further information on the nature of the anomaly. The number of
%% globally visible AS numbers increased by roughly 3,000. From the local
%% increase in AS numbers, organizations can conclude that other
%% organizations see traffic originating from the same source networks as
%% the global anomaly, because the global peak of AS numbers is not
%% bigger than their own local peak. Furthermore, port entropy hints at
%% structural differences between organizations. It is increased for 1
%% and the global view while it is decreased for 3 and 4 and remains
%% unchanged in 2.

\section{Related Work}
\label{related}
% Multi-domain Network Monitoring: http://www.perfsonar.net (it's going in production at DANTE!)

% VIFF grew out of the SIMAP project which performed the sugar beet
% auction~\cite{bogetoft2008sugarbeet} and developed the
% SMCL~\cite{nielsen2007dsp} programming language for domain-specific MPC.

Most related to our work, Roughan and Zhan~\cite{roughan2006sdd}
first proposed the use of MPC techniques for a number of applications
relating to traffic measurements, including the estimation of
global traffic volume and performance measurements~\cite{roughan2006ppp}.
In addition, the authors identified that MPC techniques can
be combined with commonly-used traffic analysis methods and tools, 
such as time-series algorithms and sketch data structures. Our work
is similar in spirit, yet it extends their work in that we introduce
new MPC protocols for event correlation, entropy, and distinct count computation and
in that we implemented these protocols in a ready-to-use library.

Data correlation systems that provide strong privacy guarantees for the
participants achieve data privacy by means of (partial) data sanitization
based on bloom filters \cite{stolfo2004wae} or cryptographic functions
\cite{lincoln2004alerts,lee2006ppi}. However, data sanitization is in general
not a lossless process and therefore imposes an unavoidable tradeoff between
data privacy and data utility.

The work presented by Chow \emph{et al.}~\cite{chow2009tpc} and Ringberg 
\emph{et al.}~\cite{ringberg2009cpp} avoid this
tradeoff by means of cryptographic data obfuscation.
Chow \emph{et al.}~proposed a two-party query computation model to perform
privacy-preserving querying of distributed databases. In
addition to the databases, their solution comprises three entities: the
randomizer, the computing engine, and  the query frontend. 
Local answers to queries are randomized by each database and the aggregate
results are de-randomized at the frontend.
%The randomizer forwards a query to each database along with a set of randomization parameters
%and sends the respective derandomization parameters to the query front end.
%Every database independently computes the query response and obfuscates the
%result with randomization parameters. The computing engine combines the
%individual obfuscated results into the final query result and sends it to the
%query front end. The query front end then uses to derandomization parameters to
%deobfuscate the final query result. Note that the entities are not considered
%to be trusted parties in the common sense but are only required not to
%collaborate.
Ringberg \emph{et al.}~present a semi-centralized solution for the
collaboration among a large number of participants in which
responsibility is divided between a proxy and a central database. In a first step the proxy obliviously blinds the
clients' input, consisting of a set of keyword/value pairs, and stores the
blinded keywords along with the non-blinded values in the central database.
On request, the database identifies the (blinded) keywords that have values
satisfying some evaluation function and forwards the matching rows to the proxy,
which then unblinds the respective keywords. Finally, the database publishes
its non-blinded data for these keywords.
%Note that in both approaches, the entities are not considered
%to be trusted parties in the common sense but are only required not to collaborate.
As opposed to these approaches, SEPIA does not depend on two central
entities but in general supports an arbitrary number of distributed privacy peers,
is provably secure, and more flexible with respect to the functions that can be
executed on the input data. 
%Two problems related to matching against private data sets are
%privacy-preserving set intersection (in which each party wants to learn
%the intersection of all private data sets) and privacy-preserving set
%matching (in which each party wants to learn whether its elements can be
%matched in any private set of the other parties). Efficient solutions
%to these problems have been proposed by Freedman et al.~\cite{freedman2004epm}
%and Sang et al.~\cite{sang2006epp}. The former solution is based on homomorphic
%encryption and balanced hashing, the latter on a threshold crypto system
%which is additive homomorphic.
The similarities and differences between our work and existing general-purpose 
MPC frameworks are discussed in Sec.~\ref{sec:eval_frameworks}.

% The authors evaluated their approach with a prototype
% implementation. Their results show that the performance scales linearly with
% the available computing resources, making it possible to improve performance
% by adding more central computing power.

\section{Conclusion}\label{sec:conclusion}

The aggregation of network security and monitoring data is crucial for
a wide variety of tasks, including collaborative network defense and
cross-sectional Internet monitoring. Unfortunately, concerns regarding
privacy prevent such collaboration from materializing. In this paper,
we investigated the practical usefulness of solutions based on secure
multiparty computation (MPC).  For this purpose, we designed optimized
MPC operations that run efficiently on voluminous input data.  We
implemented these operations in the SEPIA library along with a set of
novel protocols for event correlation and for computing multi-domain
network statistics, i.e., entropy and distinct count.  Our evaluation
results clearly demonstrate the efficiency and scalability of SEPIA in
realistic settings. With COTS hardware, near real-time operation is
practical even with 140 input providers and 9 computation
nodes. Furthermore, the basic operations of the SEPIA library are
significantly faster than those of existing MPC frameworks and can be
used as building blocks for arbitrary protocols.  We believe that our
work provides useful insights into the practical utility of MPC and
paves the way for new collaboration initiatives. Our future work
includes improving SEPIA's robustness against host failures, dealing
with malicious adversaries, and further improving performance, using,
for example, polynomial set representations.  Furthermore, in
collaboration with a major systems management vendor, we have started
a project that aims at incorporating MPC primitives into a mainstream
traffic profiling product.

\section*{Acknowledgments}

We are grateful to SWITCH for providing their traffic traces.  Also,
we want to thank Lisa Barisic and Dominik Schatzmann for their
contributions and Vassilis Zikas for assisting with MPC matters.

\bibliographystyle{abbrv}

%\small
\footnotesize
%\scriptsize
\bibliography{sepia}

\end{document}